\documentclass[11pt,a4paper]{article}
\usepackage{jheppub}
\usepackage{booktabs}

\usepackage{multirow}
\usepackage[per-mode=fraction,separate-uncertainty=true]{siunitx}
\usepackage{makecell}
\usepackage[compat=1.1.0]{tikz-feynhand}
\usetikzlibrary{shapes.misc}

\allowdisplaybreaks

\begin{document}
\title{The Singly-Charged Scalar Singlet as the\\ Origin of Neutrino Masses}

\author[a]{Tobias Felkl}
\author[b]{Juan Herrero-Garc\'{\i}a}
\author[a]{Michael A.~Schmidt}

\affiliation[a]{School of Physics, The University of New South Wales, Sydney, NSW 2052, Australia\\
Sydney Consortium for Particle Physics and Cosmology}
\affiliation[b]{Departamento de F\'isica Te\'orica and IFIC, Universidad de Valencia-CSIC, C/Catedr\'atico Jos\'e Beltr\'an, 2, E-46980 Paterna, Spain}

\emailAdd{t.felkl@unsw.edu.au}
\emailAdd{juan.herrero@ific.uv.es}
\emailAdd{m.schmidt@unsw.edu.au}

\arxivnumber{2102.09898}

\preprint{\begin{minipage}{17ex}CPPC-2021-01 \\ IFIC/21-04\end{minipage}}

\abstract{
We consider the generation of neutrino masses via a singly-charged scalar singlet. Under general assumptions we
identify two distinct structures for the neutrino mass matrix.
This yields a constraint for the antisymmetric Yukawa coupling of the singly-charged scalar singlet to two left-handed lepton doublets, irrespective of how the breaking of lepton-number conservation is achieved. The constraint disfavours large hierarchies among the Yukawa couplings. We study the implications for the phenomenology of lepton-flavour universality, measurements of the $W$-boson mass, flavour violation in the charged-lepton sector and decays of the singly-charged scalar singlet. We also discuss the parameter space that can address the Cabibbo Angle Anomaly.
}

\maketitle

%\tableofcontents
%\newpage

\section{Introduction}

The Standard Model (SM) of particle physics has been extraordinarily successful. It describes all observed fundamental particles and their gauge interactions and accounts for the masses of the charged fermions. However, the picture painted by the SM is incomplete since it predicts neutrinos to be massless.
Several neutrino-oscillation experiments including Super-Kamiokande~\cite{Fukuda:1998mi} and SNO~\cite{Ahmad:2001an,Ahmad:2002jz} established conclusive evidence that neutrinos are massive, which substantiates the need for new physics.

Introducing a Dirac mass term may be considered the most straightforward way to generate neutrino masses, however, it does not provide an explanation for their smallness.
Thus, Majorana neutrinos are generally favoured from a theoretical point of view. A Majorana mass term is generated once the SM is considered a low-energy Effective Field Theory (EFT) via the Weinberg operator~\cite{Weinberg:1979sa}, which is the lowest-dimensional non-renormalisable operator and violates the conservation of lepton number by two units. Then, neutrino masses are suppressed by the associated new-physics scale and hence are naturally small. Among the numerous ultraviolet (UV) completions of the Weinberg operator are the different seesaw mechanisms~\cite{Minkowski:1977sc,Yanagida:1979as,GellMann:1980vs,Mohapatra:1979ia,Glashow:1979nm,%
Magg:1980ut,Schechter:1980gr,Lazarides:1980nt,Wetterich:1981bx,Mohapatra:1980yp,%
Foot:1988aq} at tree level.
The first one- and two-loop neutrino mass models have been proposed more than 30 years ago~\cite{Zee:1980ai,Cheng:1980qt,WOLFENSTEIN198093,Zee:1985rj,Zee:1985id,Babu:1988ki} and in the past 20 years many more models have been designed, as detailed in various reviews on neutrino mass models~\cite{Ma:2009dk,Boucenna:2014zba,Cai:2017jrq}.

In recent years, several groups followed different avenues to systematically study neutrino mass models, based on simplicity~\cite{FileviezPerez:2009ud,Law:2013dya,Klein:2019iws}, topology~\cite{Ma:1998dn,Bonnet:2012kz,Sierra:2014rxa,Cepedello:2017eqf,Cepedello:2018rfh,Anamiati:2018cuq,Farzan:2012ev,Law:2013saa,Restrepo:2013aga}, effective operators
of the form $LLHH (H^\dagger H)^n$ with $n=0,1,\dots$~\cite{Bonnet:2009ej,Krauss:2013gy,Anamiati:2018cuq} and more generally effective operators which violate lepton number by two units ($\Delta L=2$)~\cite{Babu:2001ex,deGouvea:2007qla,deGouvea:2014lva,Angel:2012ug,Cai:2014kra,Gargalionis:2020xvt,delAguila:2012nu,Gustafsson:2020bou}.
The last option allows for an efficient classification of a large number of models and their phenomenology associated with lepton-number violation. However, processes which do not violate lepton number generally require us to resort to explicit models which are the origin of the different $\Delta L=2$ operators. There are systematic ways to use a $\Delta L = 2$ operator as a starting point to construct a UV-complete model~\cite{Babu:2001ex,Angel:2012ug,Cai:2014kra,Gargalionis:2020xvt}. A complete set of tree-level UV completions of $\Delta L=2$ operators up to dimension 11 has been constructed in~\cite{Gargalionis:2020xvt}.
The vast number of UV completions, however, make a systematic study difficult. Lastly, some of us thus proposed a classification based on simplified models~\cite{Herrero-Garcia:2019czj} and identified 20 different particles which carry lepton number and generate neutrino masses.

In this work, we focus on a singly-charged scalar singlet $h$ which transforms under the SM gauge-symmetry group $SU(3)_c\times SU(2)_L\times U(1)_Y$ as $h\sim(1,1,1)$.\footnote{Motivated by the cocktail model~\cite{Gustafsson:2012vj}, the phenomenology of the doubly-charged scalar singlet has been studied in \cite{Geib:2015tvt}.} It features in a large number of models, including the well-known Zee model~\cite{Zee:1980ai,Cheng:1980qt,WOLFENSTEIN198093} which has recently been studied in~\cite{Herrero-Garcia:2017xdu}, the Zee-Babu model~\cite{Zee:1985id,Zee:1985rj,Babu:1988ki} of which the phenomenology has been studied in~\cite{Nebot:2007bc,Ohlsson:2009vk,Herrero-Garcia:2014hfa}, the Krauss-Nasri-Trodden (KNT) model~\cite{Krauss:2002px} and a number of other models~\cite{Cai:2014kra,Cepedello:2018rfh,Ahriche:2014cda,Chen:2014ska,Cepedello:2018rfh}.
Our approach is based on the \emph{most general form
of the Majorana neutrino mass matrix
if at least one of the external neutrinos couples
via the antisymmetric Yukawa coupling $y_h$ of $h$ to two left-handed SM lepton doublets}.
We focus on the case of only one singly-charged scalar singlet which may be light, for which there are only two possible forms of the neutrino mass matrix.
The antisymmetry of the Yukawa coupling matrix $y_h$ allows us to derive model-independent constraints for the elements $y^{ij}_h$ in terms of neutrino parameters.\footnote{For the Casas-Ibarra parametrisation in seesaw models see \cite{Casas:2001sr}, and for a general parametrisation of the neutrino mass matrix see \cite{Cordero-Carrion:2018xre, Cordero-Carrion:2019qtu}.} Under the assumption that low-energy effects of new physics are dominantly governed by $h$, we then perform a phenomenology study and derive conservative bounds on parameter space which are applicable to any model of neutrino mass generation that satisfies the above criterion. We also outline generalisations of our framework to
multiple singly-charged scalar singlets.

The paper is organised as follows. In Sect.~\ref{sec:model} we discuss the structure of the neutrino mass matrix in models with a singly-charged scalar singlet and derive constraints for its Yukawa couplings. The procedure to solve the latter is elaborated on in Sect.~\ref{sec:solution}. The resulting phenomenological predictions are discussed in Sect.~\ref{sec:pheno}. In Sect.~\ref{sec:more_singly_charged} we briefly comment on the possibility of multiple singly-charged singlet scalars. In Sect.~\ref{sec:conclusions} we summarise our findings and draw a conclusion. Technical details are presented in the appendices.

\section{Singly-Charged Scalar Singlet}
\label{sec:model}
\subsection{Lagrangian}

In the following, it is assumed that the SM is extended by singly-charged scalar particle $h$ which in particular is a singlet under $SU(2)_L$. The kinetic part of the Lagrangian pertaining to $h$ is given by
\begin{align}\label{kinetic_Lagrangian}
    \mathcal{L}_{\text{kin}} = -h^*(D^\mu D_\mu + M^2_h)h
\end{align}
with the covariant derivative $D_\mu$ containing the hypercharge gauge boson in the unbroken phase. After electroweak symmetry breaking, tree-level couplings to the photon and the $Z$ boson are generated, but not to $W^\pm$ bosons. There are also a bi-quadratic coupling $|h|^2H^\dagger H$ to the SM Higgs doublet and a quartic self-coupling $|h|^4$ at tree level, however, their respective coefficients are free parameters and they are unrelated to neutrino masses. Hence, these interactions are disregarded in the following. There is no tri-linear term involving the Higgs doublet. The overall lepton sector is now enlarged to
\begin{align}\label{leptonic_Lagrangian}
    \mathcal{L}_{\text{lept}} = y^{ij}_e\bar e_iL_jH^* + y^{ij}_hL_iL_jh + \text{h.c.}
\end{align}
with the left-handed SM lepton doublet $L_i \equiv (\nu_i,\ell_i)^T$, the charge-conjugate $\bar e_i$ of the right-handed SM charged leptons and the SM lepton Yukawa coupling matrix $y_e$ which can be assumed to be diagonal, see also Sect.~\ref{sec:solution}.\footnote{Note that expanding the contraction of weak-isospin indices yields $2y^{ij}_h\nu_i\ell_jh$, hence the physical coupling matrix is given by $2y_h$.} Repeated flavour indices $i,j$ are summed over.

Besides electric charge and baryon number, this theory features another continuous global $U(1)$ symmetry that can be identified with lepton number and is conserved by the Lagrangian in Eq.~\eqref{leptonic_Lagrangian} if one assigns $+1$ unit to $L_i$, $-1$ unit to $\bar e_i$ and in particular $-2$ units to $h$. Crucial for the following analyses is the fact that the $3\times3$ Yukawa coupling matrix
\begin{align}
    y_h =
    \left(
    \begin{array}{ccc}
        0 & y^{e\mu}_h & y^{e\tau}_h \\
        -y^{e\mu}_h & 0 & y^{\mu\tau}_h \\
        -y^{e\tau}_h & -y^{\mu\tau}_h & 0 \\
    \end{array}
    \right)
\end{align}
is antisymmetric in flavour space and therefore features a non-trivial eigenvector
\begin{equation}\label{eq:ev}
    v_h = (y^{\mu\tau}_h,-y^{e\tau}_h,y^{e\mu}_h)^T
\end{equation}
with eigenvalue zero, $y_hv_h = 0$.

\subsection{Conventions for the Neutrino Sector}

Majorana masses for the active SM neutrinos are described by a symmetric complex $3\times3$ matrix $M_\nu$. In line with the conventions in \cite{Esteban:2018azc}, we relate neutrino mass eigenstates $\nu_i$ and flavour eigenstates $\nu_\alpha$ via
\begin{align}
    \nu_\alpha = \sum^3_{i=1}U_{\alpha i}\nu_i
\end{align}
with the unitary Pontecorvo-Maki-Nakagawa-Sakata (PMNS) mixing matrix $U$, and thus $m_{\text{diag}} \equiv U^TM_\nu U$. Since three generations of active neutrinos are assumed, $m_\text{diag} = \text{diag}(m_1,m_2,m_3)$ contains two or three non-vanishing eigenvalues. We have
\begin{align}
    U = PU_{23}U_{13}U_{12}U_\text{Maj}
\end{align}
with
\begin{align}
    U_{23} =
    \left(
    \begin{array}{ccc}
        1 & 0 & 0 \\
        0 & c_{23} & s_{23} \\
        0 & -s_{23} & c_{23}
    \end{array}
    \right),
    \quad
    U_{13} =
    \left(
    \begin{array}{ccc}
        c_{13} & 0 & s_{13}e^{-i\delta} \\
        0 & 1 & 0 \\
        -s_{13}e^{i\delta} & 0 & c_{13}
    \end{array}
    \right),
    \quad
    U_{12} =
    \left(
    \begin{array}{ccc}
        c_{12} & s_{12} & 0 \\
        -s_{12} & c_{12} & 0 \\
        0 & 0 & 1 \\
    \end{array}
    \right),
\end{align}
$U_{\text{Maj}} \equiv \text{diag}(e^{i\eta_1},e^{i\eta_2},1)$
and $P = \text{diag}(e^{i\alpha_1},e^{i\alpha_2},e^{i\alpha_3})$. The three phases $\alpha_k$ will eventually be removed from $U$ upon a phase redefinition of the left-handed charged leptons $\ell_i$, as described in Section~\ref{sec:solution}. $\eta_{1,2}$ are the physical Majorana phases in the case of three massive neutrinos, and $c_{ij} \equiv \cos(\theta_{ij})$ and $s_{ij} \equiv \sin(\theta_{ij})$ with the leptonic mixing angles $\theta_{12}$, $\theta_{13}$ and $\theta_{23}$.
The individual neutrino masses can be expressed in terms of the lightest neutrino mass $m_0$ and the relevant squared-mass differences $\Delta m_{ij}^2\equiv m_i^2-m_j^2$,
\begin{equation}
    m_1 = m_0, \quad m_2 = \sqrt{\Delta m_{21}^2 + m^2_0}, \quad m_3 = \sqrt{\Delta m_{31}^2 + m^2_0}
\end{equation}
in the case of Normal Ordering (NO) $m_1 < m_2 \ll m_3$, and
\begin{equation}
    m_1 = \sqrt{\left|\Delta m_{32}^2\right| - \Delta m_{21}^2 + m^2_0}, \quad m_2 = \sqrt{\left|\Delta m_{32}^2\right| + m^2_0}, \quad m_3 = m_0
\end{equation}
in the case of Inverted Ordering (IO) $m_3 \ll m_1 < m_2$ of neutrino masses. The ranges of the different parameters entering $U$ and $m_{\text{diag}}$ which are compatible with current experimental data are listed in Tab.~\ref{tab:input}.

\subsection{Neutrino Mass Matrix}

In the following, we will discuss the two possible structures for the neutrino mass matrix that are obtained in the presence of one singly-charged scalar singlet $h$.
\emph{The main assumption is that the dominant contribution to neutrino masses is generated by a diagram in which one or both of the external neutrinos couple via $y_h$.} This is schematically depicted in Fig.~\ref{fig:NeutrinoMass} where the grey blob represents unspecified physics which involves the breaking of the conservation of lepton number. Hence, we require that there are no sizeable contributions to neutrino masses which are independent of the one stemming from the singly-charged scalar singlet $h$. This scenario is naturally realised in an effective field theory (EFT) for which the grey blob represents an effective operator, but it is not limited to it. Examples are provided below when discussing the two cases.
The case of multiple singly-charged scalar singlets which generate similarly large contributions to neutrino masses is commented on in Sect.~\ref{sec:more_singly_charged}.

\subsubsection{Case I: Neutrino Masses Linear in $y_h$}
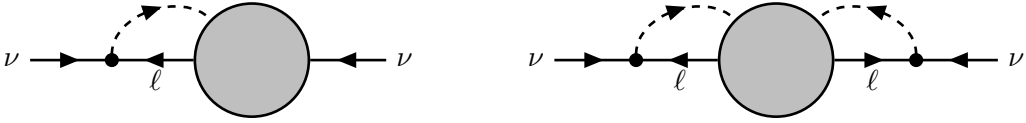
\begin{figure}[bt!]
    \centering
    \begin{tikzpicture}
    \setlength{\feynhandlinesize}{1pt}
    \tikzset{cross/.style={cross out, draw=black, thick, minimum size=2*(#1-\pgflinewidth), inner sep=0pt, outer sep=0pt},
cross/.default={3pt}}
    \begin{feynhand}
    \vertex (l) {$\nu$};
    \vertex[dot,right=of l] (vl) {};
    \vertex[grayblob,right=of vl,minimum size=1.5cm] (vr) {};
    \vertex[right=of vr] (r) {$\nu$};
    \propag[fermion] (l) to (vl);
    \propag[fermion] (r) to (vr);
    \propag[antfer] (vl) to[edge label'=$\ell$] (vr);
    \propag[chasca] (vl) to[out=90,in=135] (vr);
    \end{feynhand}
    \end{tikzpicture}
    \hspace{1cm}
    \begin{tikzpicture}
    \setlength{\feynhandlinesize}{1pt}
    \tikzset{cross/.style={cross out, draw=black, thick, minimum size=2*(#1-\pgflinewidth), inner sep=0pt, outer sep=0pt},
cross/.default={3pt}}
    \begin{feynhand}
    \vertex (l) {$\nu$};
    \vertex[dot,right=of l] (vl) {};
    \vertex[grayblob,right=of vl,minimum size=1.5cm] (v) {};
    \vertex[dot,right=of v] (vr) {};
    \vertex[right=of vr] (r) {$\nu$};
    \propag[fermion] (l) to (vl);
    \propag[fermion] (r) to (vr);
    \propag[antfer] (vl) to[edge label'=$\ell$] (v);
    \propag[antfer] (vr) to[edge label=$\ell$] (v);
    \propag[chasca] (vl) to[out=90,in=135] (v);
    \propag[chasca] (vr) to[out=90,in=45] (v);
    \end{feynhand}
    \end{tikzpicture}

    \caption{Self-energy diagram responsible for the generation of neutrino masses via a singly-charged scalar singlet: linear case (left) and quadratic case (right). The grey blob represents all other interactions which contribute to the diagram. It could be one effective vertex or a sub-diagram consisting of multiple vertices and propagators. There are at least two insertions of the Higgs vacuum expectation value somewhere in the diagram which are not explicitly shown.}
    \label{fig:NeutrinoMass}
\end{figure}
If the main contribution to neutrino masses is generated by a diagram in which only one of the external neutrinos couples via $y_h$, as schematically shown on the left in Fig.~\ref{fig:NeutrinoMass}, neutrino masses are approximately given by
\begin{align}\label{amplitude_one_singly_charged}
    U^*m_{\text{diag}}U^\dagger = M_\nu = Xy_h - y_hX^T.
\end{align}
Here, the coupling matrix $X$ contains the information about the rest of the loop structure, that is, particle masses, couplings, loop factors and further unknown parameters. It is stressed again that the main assumption that there are no sizeable contributions to neutrino masses which cannot be parametrised as above is essential for what follows.
Multiplying Eq.~(\ref{amplitude_one_singly_charged}) by $v_h$ defined in Eq.~\eqref{eq:ev} from the left- and the right-hand side, one obtains
\begin{align}\label{constraint_eigenvector}
    v^T_hU^*m_{\text{diag}}U^\dagger v_h = 0\,,
\end{align}
which we identify as a \emph{necessary condition} for neutrino masses being correctly explained by $h$. Eq.~\eqref{constraint_eigenvector} is very predictive in the sense that is does not involve $X$ and hence the mechanism of the breaking of lepton-number conservation does not have to be specified. Instead, we maintain a model-independent approach throughout the analysis and do not explicitly construct the neutrino mass matrix. Treating the elements of $X$ as essentially free parameters also implies that in general the determinant of $M_\nu$ does not vanish and hence all three active neutrinos are massive.\footnote{Linear combinations of the elements $X^{ij}$ are constrained in the sense that Eq.~(\ref{amplitude_one_singly_charged}) has to be satisfied, however, this does not uniquely determine the $X^{ij}$ in terms of the $y^{kl}_h$ since $M_\nu$ is symmetric, whereas $X$ can be a general matrix.} Nevertheless, one may impose $\det(M_\nu) = 0$ as a further condition which then necessarily also involves the elements of $X$. In this case, the smallest neutrino mass and one of the Majorana phases vanish, the consequences of which will be briefly commented on in Sect.~\ref{sec:moc}. See App.~\ref{sec:nmc} for the expression in Eq.~\eqref{constraint_eigenvector} explicitly written out.

In case $X$ is generated by some heavy new physics, one may use EFT to parametrise its effect. As an example, let us consider the non-renormalisable dimension-5 operator \cite{Herrero-Garcia:2019czj}
\begin{align}\label{effective_dim5_operator_linear}
    \frac{c^{ij}}{\Lambda}h^*\bar e_iL_jH + \text{h.c.}
\end{align}
which violates lepton number by two units. Here, the lepton-number breaking scale $\Lambda$ is assumed to be much larger than $M_h$, and $c$ is a general complex $3\times 3$ matrix. Then, neutrino masses are generated at one-loop level and can be approximately written as
\begin{align}\label{amplitude-singly-charged}
    M_\nu \propto \frac{v^2}{(4\pi)^2\Lambda}\left(c\,y_e\,y_h - y_h y_e\,c^T\right).
\end{align}
Hence, in this case
\begin{equation}
X \approx \frac{c\,y_e}{(4\pi)^2\Lambda} v^2\,.
\end{equation}
There is in principle an infinite number of potential realisations of this effective description of neutrino masses in terms of concrete models. Among them, several simple examples in which neutrino masses are generated at three-loop level are discussed in \cite{Cepedello:2018rfh}.\footnote{They are dubbed `Model 3' and `Model 4' therein.
Another possibility mentioned is to take $h$ as accompanied by the scalar doublet $\sim(1,2,3/2)$ and generate neutrino masses at two-loop level. This can be seen as a modification of the Zee model in the sense that one of the loops generates the tri-linear term $HHh + \text{h.c.}$ which vanishes at tree level.}
In addition, the constraint in Eq.~\eqref{constraint_eigenvector} also applies to some of the minimal UV completions of lepton-number violating effective operators discussed in \cite{Cai:2014kra}. Still, the most prominent realisation of the general structure in Eq.~\eqref{amplitude_one_singly_charged} is given by the Zee model and its variants \cite{Zee:1980ai,Cheng:1980qt,WOLFENSTEIN198093,Herrero-Garcia:2017xdu,Cai:2017jrq,He:2011hs,Matsui:2021khj}. Here, the SM particle content is enlarged by $h$ and a new Higgs doublet $\Phi\sim(1,2,1/2)$ which in particular allows for a tri-linear term $H\Phi h^* + \text{h.c.}$ at tree level which violates lepton number. Then,
\begin{equation}
    X = y'_e\,m_e\,\frac{\sin(2\varphi)}{16 \pi^2}\, \log \left(\frac{M^2_{h^+_2}}{M^2_{h^+_1}} \right)\,,
\end{equation}
with  $m_e$ the SM charged-lepton masses, $\varphi$ the angle parametrising the mixing of the singly-charged scalar mass eigenstates $h^+_{1,2}$ with masses $M_{h^+_{1,2}}$, and $y'_e$ the Yukawa coupling of $\Phi$ (in the so-called Higgs basis) to the SM leptons. Together with the tri-linear term, the latter generates the effective operator in Eq.~\eqref{effective_dim5_operator_linear} at tree level when the second Higgs doublet $\Phi$ is integrated out.

\subsubsection{Case II: Neutrino Masses Quadratic in $y_h$}

If neutrino masses are dominantly generated by a diagram in which both external neutrinos couple via $y_h$, respectively, as shown on the right in Fig.~\ref{fig:NeutrinoMass}, one obtains
\begin{align}\label{amplitude_one_singly_charged_symmetric}
    U^*m_{\text{diag}}U^\dagger = M_\nu = y_h\,S\,y_h\,,
\end{align}
where $S$ is a complex symmetric matrix. This can be considered as a special realisation of the linear case (Case I) with $X = y_hS'$, where $S'$ is a general complex matrix and thus $S \equiv S' + S'^T$. Still, this identification is trivial if the main contribution to neutrino masses is inherently flavour-symmetric.
The lightest neutrino will be massless at this order, because the determinant of $M_\nu$ vanishes by construction due to $y_h$ being antisymmetric. Also, this implies that there is only one physical Majorana phase.\footnote{We choose $\eta_1 = 0$ in the quadratic case which matches the convention in \cite{Nebot:2007bc}.}

As in the linear case, the relevant assumption is that the model under consideration does not generate any sizeable contribution to neutrino masses which is not given by the structure in Eq.~\eqref{amplitude_one_singly_charged_symmetric}.
Then, one identifies the condition
\begin{align}\label{constraint_eigenvector_special}
    m_{\text{diag}}U^\dagger v_h = 0
\end{align}
which trivially implies the one in Eq.~\eqref{constraint_eigenvector}, but the converse statement is not true in general. Explicitly, Eq.~\eqref{constraint_eigenvector_special} yields the two relations
\begin{align}
    \frac{y^{e\tau}_h}{y^{\mu\tau}_h} & = \tan(\theta_{12})\frac{\cos(\theta_{23})}{\cos(\theta_{13})} + \tan(\theta_{13})\sin(\theta_{23})e^{i\delta}, \\
    \frac{y^{e\mu}_h}{y^{\mu\tau}_h} & = \tan(\theta_{12})\frac{\sin(\theta_{23})}{\cos(\theta_{13})} - \tan(\theta_{13})\cos(\theta_{23})e^{i\delta}
\end{align}
in the case of NO and
\begin{align}
    \frac{y^{e\tau}_h}{y^{\mu\tau}_h} & = -\frac{\sin(\theta_{23})}{\tan(\theta_{13})}e^{i\delta}, \\
    \frac{y^{e\mu}_h}{y^{\mu\tau}_h} & = \frac{\cos(\theta_{23})}{\tan(\theta_{13})}e^{i\delta}
\end{align}
for IO.\footnote{Eq.~\eqref{constraint_eigenvector_special} formally implies three equations, but one of them is trivially satisfied due to $\det(M_\nu) = 0$. Also, the expressions differ from the ones in \cite{Nebot:2007bc} by a complex conjugation as per how the PMNS matrix $U$ is defined.} Note that these relations only depend on the leptonic mixing angles and the Dirac CP phase and are independent of the Majorana phases and individual neutrino masses, and thus they are more constraining than the one in Eq.~\eqref{constraint_eigenvector}.

For a concrete example for $S$ in terms of an EFT, one may consider the non-renormalisable dimension-5 operator
\begin{align}\label{effective_dim5_operator_quadratic}
    \frac{d^{ij}}{\Lambda}(h^*)^2\bar e_i\bar e_j + \text{h.c.}\,
\end{align}
which violates lepton number by two units. Here, the lepton-number breaking scale $\Lambda$ is assumed to be much larger than $M_h$, and $d$ is a complex symmetric $3\times 3$ matrix. Then, neutrino masses are generated at two-loop level and can be approximately written as
\begin{align}\label{amplitude-singly-charged-symmetric}
    M_\nu \propto \frac{v^2}{(4\pi)^4\Lambda}y_h\,y_e\,d\,y_e\,y_h\,.
\end{align}
Hence, in this case
\begin{equation}
S \approx \frac{y_edy_e}{(4\pi)^4\Lambda} v^2\,.
\end{equation}
The constraint in Eq.~\eqref{constraint_eigenvector_special} has been previously discussed~\cite{Babu:2002uu,Nebot:2007bc,Ohlsson:2009vk,Herrero-Garcia:2014hfa} in the context of the Zee-Babu model~\cite{Zee:1985rj,Zee:1985id,Babu:1988ki}.
Here, the SM particle content gets enlarged by $h$ and a doubly-charged scalar singlet $k\sim(1,1,2)$ with mass $M_k$ which in particular allows for a tri-linear term $\mu h^2k^* + \text{h.c.}$ at tree level which violates lepton number. Then, neutrino masses are generated at two-loop level and one may write \cite{Nebot:2007bc,Herrero-Garcia:2014hfa,McDonald:2003zj}
\begin{equation} \label{eq:SZB}
S = 16 \, m_e\,y_k\,m_e\, \mu\, F\left(\frac{M^2_{k}}{M^2_{h}} \right)\,,
\end{equation}
with $y_k$ the symmetric Yukawa coupling matrix  of $k$ to right-handed SM leptons and $F$ a loop function. The effective operator in Eq.~\eqref{effective_dim5_operator_quadratic} is induced at tree level when $k$ is integrated out. However, the constraint also applies to the KNT model \cite{Krauss:2002px} which features a second singly-charged scalar singlet $\sim(1,1,1)$ and a fermionic singlet $\sim(1,1,0)$ both of which are charged  under a $Z_2$ symmetry, as well as to some variants of it discussed in~\cite{Ahriche:2014cda,Chen:2014ska,Cepedello:2018rfh} or to the extension of the Zee-Babu model by another heavy singly-charged scalar singlet, see App.~\ref{sec:gzbm}. Analogous to $X$ in the linear case, the constraint in Eq.~\eqref{constraint_eigenvector_special} does not involve $S$ itself and hence is independent of the details of the breaking of lepton-number conservation.

\section{Solving the Neutrino-Mass Constraint}
\label{sec:solution}

In this section, the procedure of solving the constraint in Eq.~(\ref{constraint_eigenvector}) is elaborated on. Both the real part and the imaginary part of $v^T_hU^*m_{\text{diag}}U^\dagger v_h$ have to identically vanish which yields two real conditions. We decompose the couplings into their respective magnitudes and phases and use the constraint to determine two of the $|y^{ij}_h|$ in terms of the third one, the phases and the active-neutrino parameters which enter $m_{\text{diag}}$ and $U$. This amounts to finding the roots of a single expression that is quartic in two of the $y^{ij}_h$ since both the real and imaginary part of $v^T_hU^*m_{\text{diag}}U^\dagger v_h$ can be taken as quadratic in either of the couplings $y^{ij}_h$. Therefore, the constraint is numerically solved and one can obtain up to four solutions. In the quadratic case, Eq.~\eqref{constraint_eigenvector_special} implies four real conditions which then also determine two phases of the Yukawa couplings $y^{ij}_h$ in terms of neutrino data.

The smallest neutrino mass $m_0$ can be arbitrarily small or even zero, whereas upper bounds arise from cosmological surveys as well as experimental searches for tritium beta decay and neutrino-less double beta decay. The cosmological bound is the strongest one and, while model-dependent, it is assumed to apply in the scenario under consideration since no new physics is introduced below the electroweak scale.
The latest results published by the Planck Collaboration in 2018 \cite{Aghanim:2018eyx} comprise the upper bound $m_1 + m_2 + m_3 \le \SI{0.12}{\electronvolt}$ which implies $m_0 = m_1 \lesssim \SI{30}{\milli\electronvolt}$ for NO and $m_0 = m_3 \lesssim \SI{15}{\milli\electronvolt}$ for IO. $|y^{ij}_h|$ can in principle also be arbitrarily small, whereas $|y^{ij}_h|\lesssim 2\pi$ due to perturbativity constraints with the normalisation of the Yukawa coupling taken into account.\footnote{The constraint $|y^{ij}_h| \lesssim 2\pi$ may for instance be derived from requiring that the one-loop correction to the physical coupling $2y^{ij}_h$ remains smaller than $2y^{ij}_h$ itself.} Notwithstanding, both the coupling magnitude assigned a value and the magnitudes obtained as solutions to Eq.~(\ref{constraint_eigenvector}) are required to satisfy $|y^{ij}_h| > \num{E-4}$ in order to limit the orders of magnitude sampled over.

The flavour observables discussed in the following section also depend on the mass $M_h$ which is not constrained by Eq.~\eqref{constraint_eigenvector}. At the LHC the singly-charged scalar singlet is dominantly produced in pairs via the Drell-Yan process and decays to a charged lepton and a neutrino. Hence, for decays to electrons and muons the signal is a pair of oppositely-charged leptons with missing energy. The branching ratio of the tau-lepton channel cannot exceed 50\%, as it is shown in Sec.~\ref{sec:dcss}. A model-independent lower bound $M_h \gtrsim \SI{200}{\giga\electronvolt}$ has recently been derived in \cite{Crivellin:2020klg} from the reinterpretation of a collider search for smuons and selectrons \cite{Aad:2019vnb}. Depending on the relative magnitudes $|y^{ij}_h|$, the constraint is actually slightly more stringent. Hence, we require $M_h \ge \SI{350}{\giga\electronvolt}$ to safely operate beyond any mass region potentially excluded. This is consistent with the earlier analysis in \cite{Cao:2017ffm}. The assumed upper bound $M_h\le\SI{100}{\tera\electronvolt}$ arises from an order-of-magnitude estimate based on requiring the absence of unnaturally large corrections to the SM Higgs-boson mass \cite{Herrero-Garcia:2019czj}.

Furthermore, a careful determination of the physical phases in the theory is in order. Before electroweak symmetry breaking, unitary basis transformations applied to $\bar e_i$ and $L_i$ can be used to diagonalise $y_e$ with real and positive eigenvalues, and the phases in $y_h$ can be eliminated upon redefinitions of $\bar e_i$ and $L_i$. After electroweak symmetry breaking, the charged-lepton masses are already diagonal by construction, and the neutrino mass matrix is diagonalised via the PMNS matrix $U$. Then, three phases in $U$ can be eliminated via redefining the left-handed charged leptons $\ell_i$ which reintroduces three phases in $y_h$. One of these can be set to zero upon exploiting the phase freedom of $h$. Therefore, $y^{\mu\tau}_h$ is taken real while $\arg(y^{e\mu}_h)$ and $\arg(y^{e\tau}_h)$ are randomly sampled over. As a side note, the presence of complex couplings indicates that the singly-charged scalar singlet, accompanied by a source of lepton-number violation, will in general contribute to leptonic electric dipole moments. However, as these are linked to the violation of lepton-number conservation and hence no strong constraints are to be expected, electric dipole moments are not explored further. As of yet, the physical Majorana phases $\eta_{1,2}$ are completely unconstrained experimentally and hence also randomly sampled over. Note that the ranges of $\eta_{1,2}$ can be restricted to $[0,\pi]$ without loss of generality since the sign of the Majorana field is unphysical.
\begin{table}[tb!]\centering
    \setlength{\tabcolsep}{10pt}
    \renewcommand{\arraystretch}{1.5}
%\begin{ruledtabular}
        \begin{tabular}{c c c c c c}\toprule
            $m_e\,[\si{\kilo\electronvolt}]$ & $m_\mu\,[\si{\mega\electronvolt}]$ & $m_\tau\,[\si{\giga\electronvolt}]$ &
            $G_F\,[\si{\per\giga\electronvolt\squared}]$ &
            $\alpha^{-1}_{\text{EM}}$ & $M_Z\,[\si{\giga\electronvolt}]$ \\
            \midrule
            \num{510.9989} & \num{105.6584} & \num{1.777} &
            $1.16638\times 10^{-5}$ & 137.035999 & \num{91.1535}
            \\\bottomrule
            \end{tabular}
            \label{leptonic_input_parameters}
 %       \end{ruledtabular}
   \vspace{5ex}

        %\begin{ruledtabular}
        \begin{tabular}{c|c c c}\toprule
            & $\Delta m^2_{3l}$\,[\SI{E-3}{\electronvolt\squared}] & $\Delta m^2_{21}$\,[\SI{E-5}{\electronvolt\squared}] & $\delta$\;[\si{\radian}] \\
            \midrule
            NO & \num{2.517\pm0.026} & \num{7.42\pm0.20}
            & \num{3.44\pm0.42}
            \\
            IO & \num{-2.498\pm0.028} & \num{7.42\pm0.20}
            & \num{4.92\pm0.45}
            \\
            \midrule
            & $\sin^2(\theta_{12})$ & $\sin^2(\theta_{13})$ & $\sin^2(\theta_{23})$ \\
            \midrule
            NO & \num{0.304\pm0.012}{} & $0.02219\pm0.00062$ & \num{0.573\pm0.016}
            \\
            IO & \num{0.304\pm0.012}{} & $0.02238\pm0.00062$ & \num{0.575\pm0.016}
            \\\bottomrule
        \end{tabular}
    %\end{ruledtabular}
    \vspace{5ex}

%\begin{ruledtabular}
        \resizebox{\textwidth}{!}{%
        \begin{tabular}{c|c c c c c c}\toprule
            & $|y^{ij}_h|$ & $\arg(y^{ek}_h)$ & $\arg(y^{\mu\tau}_h)$ & $m_0$\;[\si{\milli\electronvolt}] & $\eta_{1,2}$\;[\si{\radian}] & $M_h\;[\si{\giga\electronvolt}]$ \\
            \midrule
            Prior & Log-Flat & Flat & Fixed & Log-Flat & Flat & Log-Flat \\
            Range & $[\num{E-4},2\pi]$ & $[0,2\pi]$ & $0$ & \makecell{$[\num{E-4},\;\num{30}]$ (NO) \\ [-0.25ex]
            $[\num{E-4},\;\num{15}]$ (IO)} & $[0,\pi]$ & [\num{350},\;\num{E5}]
            \\\bottomrule
            \end{tabular}
            }
 %       \end{ruledtabular}
   \caption{
   The table at the top contains the experimental values for input parameters taken from \cite{Zyla:2020zbs,Freitas:2020kcn}; the one in the centre
  summarises the experimental values for leptonic mixing parameters and neutrino-mass-squared differences taken from NuFIT 5.0~\cite{Esteban:2020cvm}. $\Delta m^2_{31} > 0$ for NO, and $\Delta m^2_{32} < 0$ for IO. Priors and the ranges sampled over in the numerical scan are given in the bottom table. The flavour indices for the antisymmetric Yukawa couplings are $i,j=e,\,\mu,\,\tau$, and $k=\mu,\,\tau$.}
    \label{tab:input}
\end{table}

The leptonic mixing parameters and neutrino-mass-squared differences which have been constrained by experiments are assigned pseudo-random variates from normal distributions of which the respective mean values and standard deviations are taken from the latest fit results provided by the NuFIT collaboration~\cite{Esteban:2020cvm}.
Symmetric distributions are assumed for simplicity. The numerical values used for the charged-lepton masses $(m_e,m_\mu,m_\tau)$ and the electroweak input parameters ($G_F,\alpha_{\text{EM}},M_Z$) are summarised in Tab.~\ref{tab:input}. For the numerical scan we generated approximately \num{5E5} sample points for each neutrino-mass ordering.

%%%%%%%%%%%%%%%%%%%%%%%%%%%%%%%%%%%%%%%%%%%%%%%%%%%%%%%%%%%%
%%%%%%%%%%%%%%%%%%%%%%%%%%%%%%%%%%%%%%%%%%%%%%%%%%%%%%%%%%%%
\section{Phenomenology}
\label{sec:pheno}

In the following, the contributions of $h$ to various flavour observables are presented under the assumption that the couplings $y^{ij}_h$ satisfy the constraints in Eq.~\eqref{constraint_eigenvector} or Eq.~\eqref{constraint_eigenvector_special} and hence neutrino masses are dominantly generated by the singly-charged scalar singlet. The considered observables together with the current experimental bound, prospected sensitivities for future experiments as well as the maximum contribution found in the numerical scan are summarised in Tab.~\ref{tab:observables}. Note that the bounds on several observables can be (nearly) saturated. Large tuning of the ratios of coupling magnitudes $|y^{ij}_h|$ is necessarily absent due to the constraints in Eq.~\eqref{constraint_eigenvector} and Eq.~\eqref{constraint_eigenvector_special}, see Sect.~\ref{sec:moc}, thus in particular the bounds on $\mu\to e\gamma$ and $\mu\to 3e$ cannot be evaded and hence they efficiently shape the available parameter space.

We assume that further new particles are weakly coupled or heavy enough not to generate sizeable contributions to any of the observables under consideration. In particular, particles which induce flavour-changing decays of charged leptons at tree level have to be sufficiently decoupled, as the singly-charged scalar singlet generates these processes at one-loop level. Significant destructive interference or fine-tuned cancellations are taken as absent. Succinctly, we assume that low-energy effects of new physics are dominantly governed by $h$. In that sense, the bounds on parameter space which is compatible with neutrino masses as discussed in the following are conservative.
\begin{table*}[bt!]
   \centering
   {  \small
   \setlength{\tabcolsep}{10pt}
    \renewcommand{\arraystretch}{1.5}
  %  \begin{ruledtabular}
        \begin{tabular}{c|cc}\toprule & \multicolumn{2}{c}{Experimental Data} \\
            Observable & Current Bound & Future Sensitivity \\
            \midrule
            $\text{Br}(\mu\to e\gamma)$ & \num{4.2E-13} (90\% CL) \cite{TheMEG:2016wtm} & \num{6E-14} \cite{Baldini:2018nnn} \\
            $\text{Br}(\tau\to e\gamma)$ & \num{3.3E-8} (90\% CL) \cite{Aubert:2009ag} & \num{3E-9} \cite{Kou:2018nap} \\
            $\text{Br}(\tau\to \mu\gamma)$ & \num{4.4E-8} (90\% CL) \cite{Aubert:2009ag} & \num{E-9} \cite{Kou:2018nap} \\
            $\text{Br}(\mu\to 3e)$ & \num{E-12} (90\% CL) \cite{Bellgardt:1987du} & \num{E-16} \cite{Arndt:2020obb} \\ $\text{Br}(\tau\to 3e)$ & \num{2.7E-8} (90\% CL) \cite{Hayasaka:2010np} & \num{4.3E-10} \cite{Kou:2018nap} \\
            $\text{Br}(\tau\to 3\mu)$ & \num{2.1E-8} (90\% CL) \cite{Hayasaka:2010np} & \num{3.3E-10} \cite{Kou:2018nap} \\
            $|g_\mu/g_e|$  & [\num{0.9986},\;\num{1.0050}] $(2\sigma)$ \cite{Gersabeck:2020ywo} \\
            $|g_\tau/g_\mu|$  & [\num{0.9981},\;\num{1.0041}] $(2\sigma)$ \cite{Gersabeck:2020ywo} \\ $|g_\tau/g_e|$ & \makecell{[\num{1.0000},\;\num{1.0060}] $(2\sigma)$ \cite{Gersabeck:2020ywo} \\ [-0.25ex]
            [\num{0.9985},\;\num{1.0075}] $(3\sigma)$ \cite{Gersabeck:2020ywo}} \\
            $|\delta M_W| [\si{\giga\electronvolt}]$ &
            \num{0.018} $(3\sigma)$
            \cite{Zyla:2020zbs} \\\bottomrule
        \end{tabular}
   %     \end{ruledtabular}
   \vspace{5ex}

        \begin{tabular}{c|cc|cc}\toprule
        & \multicolumn{4}{c}{Numerical Analysis} \\
        & \multicolumn{2}{c}{Linear Case} & \multicolumn{2}{c}{Quadratic Case}. \\
            Observable & NO & IO & NO & IO \\
            \midrule
            $\text{Br}(\mu\to e\gamma)$ & \num{4.2E-13} & \num{4.2E-13} & \num{4.2E-13} & \num{4.2E-13} \\
            $\text{Br}(\tau\to e\gamma)$ & \num{6.4E-11} & \num{4.9E-11} & \num{3.1E-13} & \num{6.8E-14} \\
            $\text{Br}(\tau\to \mu\gamma)$ & \num{1.6E-11} & \num{1.6E-11} & \num{2.9E-14} & \num{1.5E-12} \\
            $\text{Br}(\mu\to 3e)$ & \num{E-12} & \num{E-12} & \num{E-12} & \num{E-12} \\ $\text{Br}(\tau\to 3e)$ & \num{6.6E-9} & \num{1.3E-8} & \num{7.7E-13} & \num{1.6E-13} \\
            $\text{Br}(\tau\to 3\mu)$ & \num{3.0E-9} & \num{1.2E-8} & \num{6.1E-13} & \num{8.8E-13} \\
            $|g_\mu/g_e|$ & 1.0050 & 1.0047 & 1.0002 & 1.0000 \\
            $|g_\tau/g_\mu|$ & 1.0009 & 1.0014 & 1.0000 & 1.0001 \\
            $|g_\tau/g_e|$ & 1.0048 & 1.0043 & 1.0002 & 1.0000 \\
            $|\delta M_W| [\si{\giga\electronvolt}]$ & \num{0.018} & \num{0.018} & \num{0.002} & \num{0.007} \\\bottomrule
        \end{tabular}
   }
    \caption{
    The upper table contains the current experimental bounds on and future sensitivities to the relevant observables. The lower table shows the respective maximum contribution found in the scan in the linear case and the quadratic case for either neutrino-mass ordering.
    \label{tab:observables}}
\end{table*}

\subsection{Effective Description of Low-Energy Phenomenology at Tree Level}

As derived in App.~\ref{sec:effOp}, the Wilson coefficient of the effective dimension-6 four-lepton operator
\begin{align}
    \mathcal{O}_{LL,ijkl} \equiv L^{\dagger\alpha}_i\bar\sigma^\mu L_{j\alpha}L^{\dagger\beta}_k\bar\sigma_\mu L_{l\beta}
\end{align}
receives a contribution at tree level from integrating out the singly-charged scalar singlet $h$:\footnote{See \cite{Bilenky:1993bt} for integrating out $h$ at one-loop level.}
\begin{align}\label{eq:CLL}
    C^{ijkl}_{LL} = \frac{(y^{ik}_h)^*y^{jl}_h}{M^2_h}\,.
    \;
\end{align}
In the low-energy effective theory, this leads to the neutral-current Lagrangian
\begin{align}\label{eqn:ncl}
    \mathcal{L}^{\text{NSI}}_{d=6} = -2\sqrt{2}G_F\epsilon^{kl}_{ij}\left(\nu^\dagger_i\bar\sigma^\mu \nu_j\right)\left(\ell^\dagger_k\bar\sigma_\mu \ell_l\right)\,,
\end{align}
(see also \cite{Herrero-Garcia:2014hfa}) with the Wilson coefficients
\begin{align}\label{eq:NSI}
    \epsilon_{ij}^{kl} \equiv -\frac{1}{2\sqrt{2}G_F} \left(C_{LL}^{ijkl}+C_{LL}^{klij}\right)
    = -\frac{1}{\sqrt{2}G_F}\frac{(y_h^{ik})^*y_h^{jl}}{M_h^2}
\end{align}
which are commonly called non-standard interaction (NSI) parameters.
They are antisymmetric under the exchange of an upper index and the corresponding lower index, $\epsilon_{ij}^{kl} = -\epsilon_{il}^{kj} = -\epsilon_{kj}^{il}$, and their complex conjugates are obtained via swapping the upper and lower indices among themselves: $\epsilon_{ij}^{kl}=(\epsilon_{ji}^{lk})^*$.
Note that there are no effective operators with four neutrinos or four charged leptons due to the antisymmetry of $y_h$ and thus in particular no tree-level contributions to flavour-violating charged-lepton decays.

%%%%%%%%%%%%%%%%%%%%%%%%%%%%%%%%%%%%%%%%%%%%%%%%%
\subsubsection*{Fermi Constant and CKM Matrix}

Singly-charged scalar singlets affect the partial decay widths $\Gamma_{a\to b}$ associated to the different leptonic channels $\ell^-_a\to \ell^-_b \nu_a \bar\nu_b$  and $\ell^+_a\to \ell^+_b \bar\nu_a \nu_b$ \cite{Pich:2013lsa,Nebot:2007bc,Crivellin:2020klg} and hence in particular modify the extraction of the Fermi constant $G_F$ from measurements of the muon lifetime. In the framework of treating the SM as an effective field theory (SMEFT), one defines (see e.g. \cite{Berthier:2015oma})
\begin{align}
    G_F = G^{\text{SM}}_F - \frac{\sqrt{2}}{4}\left(C^{\mu ee\mu}_{LL} + C^{e\mu\mu e}_{LL}\right)
\end{align}
with the Wilson coefficient $C^{ijkl}_{LL}$ given in Eq.~\eqref{eq:CLL} and $G_F^{\rm SM}$ denotes the Fermi constant in the SM.\footnote{Additional contributions from other operators to $G_F$ are omitted.} Hence,
\begin{align}
    G_F = G^{\text{SM}}_F - \frac{\sqrt{2}}{4}\left(\frac{(y^{e\mu}_h)^*y^{\mu e}_h}{M^2_h} + \frac{(y^{\mu e}_h)^*y^{e\mu}_h}{M^2_h}\right) = G^{\text{SM}}_F + \frac{1}{\sqrt{2}}\frac{|y^{e\mu}_h|^2}{M^2_h}\,,
\end{align}
where we have used the antisymmetry of the Yukawa coupling matrix $y_h$. Equivalently, we can express it as $G_F = G^{\text{SM}}_F + \sqrt{2}G_F\delta G_F$ with
\begin{align}
    \delta G_F = \frac{1}{2G_F}\frac{|y^{e\mu}_h|^2}{M^2_h} \equiv - \frac{\epsilon^{\mu\mu}_{ee}}{\sqrt{2}}.
\end{align}
Another observable which has recently attracted attention (see for instance \cite{Alok:2020jod,Coutinho:2019aiy,Kirk:2020wdk,Crivellin:2020ebi,Capdevila:2020rrl,Endo:2020tkb,Crivellin:2020lzu,Belfatto:2019swo}) and is of interest for the scenario under consideration is the sum of the squares of the absolute values of the first-row elements of the Cabibbo-Kobayashi-Maskawa (CKM) matrix:
\begin{align}\label{eq:ckm}
    \sum_{i}|V_{ui}|^2 = |V_{ud}|^2 + |V_{us}|^2 + |V_{ub}|^2\,.
\end{align}
The magnitude of the element $V_{us}$ can be extracted directly from kaon and tau decays \cite{Aoki:2019cca,Amhis:2019ckw}, and indirectly via $|V_{ud}|$ from nuclear beta decays (see for instance \cite{Seng:2018qru,Seng:2020wjq} for recent theoretical progress) and the assumption of the sum in Eq.~(\ref{eq:ckm}) being equal to one which in the SM is a built-in consequence of unitarity.\footnote{The magnitude of $V_{ub}$ is negligibly small in this context.} The fact that there is significant tension between the results is referred to as the Cabibbo Angle Anomaly (CAA). The discrepancy between the ``true" value of $|V_{us}|$ and the one obtained from beta decays and CKM unitarity in the SM
can be explained via new contributions to muon decay and subsequently the Fermi constant \cite{Crivellin:2020klg}.

%%%%%%%%%%%%%%%%%%%%%%%%%%%%%%%%%%%%%%%%%%%%%%%%%%%%%%%%%%%%

\subsubsection*{Universality of Leptonic Gauge Couplings}

One defines the lepton-flavour universality ratios via the ``effective Fermi constants" $G_{ab} \sim g_ag_b$ associated to the different leptonic channels:
\cite{Nebot:2007bc,Pich:2013lsa}
\begin{align}
    \sqrt[4]{\frac{\Gamma_{\tau\to\mu}}{\Gamma_{\tau\to e}}} & \propto \frac{G_{\tau\mu}}{G_{\tau e}} = \frac{g_\mu}{g_e} \approx 1 + \frac{1}{\sqrt{2}G_F}\frac{|y^{\mu\tau}_h|^2 - |y^{e\tau}_h|^2}{M^2_h} \equiv 1 +\epsilon_{ee}^{\tau\tau}- \epsilon_{\mu\mu}^{\tau\tau}, \\
    \sqrt[4]{\frac{\Gamma_{\tau\to \mu}}{\Gamma_{\mu\to e}}} & \propto \frac{G_{\tau\mu}}{G_{\mu e}} = \frac{g_\tau}{g_e} \approx 1 + \frac{1}{\sqrt{2}G_F}\frac{|y^{\mu\tau}_h|^2 - |y^{e\mu}_h|^2}{M^2_h} \equiv 1+\epsilon_{ee}^{\mu\mu}-\epsilon_{\mu\mu}^{\tau\tau} , \\
    \sqrt[4]{\frac{\Gamma_{\tau\to e}}{\Gamma_{\mu\to e}}} & \propto \frac{G_{\tau e}}{G_{\mu e}} = \frac{g_\tau}{g_\mu} \approx 1 + \frac{1}{\sqrt{2}G_F}\frac{|y^{e\tau}_h|^2 - |y^{e\mu}_h|^2}{M^2_h} \equiv 1+ \epsilon_{ee}^{\mu\mu} -\epsilon_{ee}^{\tau\tau}\,.
\end{align}
The experimental best-fit values of all three universality ratios are currently larger than one, $|g_\mu/g_e| = \num{1.0018\pm0.0032}$, $|g_\tau/g_e| = \num{1.0030\pm0.0030}$, $|g_\tau/g_\mu| = \num{1.0011\pm0.0030}$ with errors given at $2\sigma$~\cite{Gersabeck:2020ywo}. In particular, the channel $\tau\to\mu$ appears to receive sizeable contributions from new physics.

In \cite{Crivellin:2020klg} it has been shown that the deviations of $g_\mu/g_e$ and $g_\tau/g_e$ from one and the CAA, which will be collectively referred to as the ``flavour anomalies" henceforth, can be simultaneously explained with a singly-charged scalar singlet. Adopting the results for the best-fit regions and using the terminology as in \cite{Crivellin:2020klg}, for simplicity we take the anomalies to be explained if both $\delta(\mu\to e\nu\nu)\in[0.0005,0.0008]$ and $\delta(\tau\to \mu\nu\nu)\in[0.0016,0.004]$ are satisfied\footnote{These ranges are located within the region preferred at $1\sigma$ as presented in \cite{Crivellin:2020klg}. We refrain from parametrising its elliptic shape.}, with
\begin{align} \label{eq:CAAeq}
    \delta(\ell_i\to \ell_j\nu\nu) \equiv \frac{1}{\sqrt{2}G_F}\frac{|y^{ij}_h|^2}{M^2_h} = -\epsilon_{ii}^{jj}.
\end{align}
This immediately implies an upper bound $M_h \lesssim \SI{39}{\tera\electronvolt}$ if $h$ explains the flavour anomalies, given that perturbativity constraints require $|y^{ij}_h| \lesssim 2\pi$.
The experimental values used in \cite{Crivellin:2020klg} are taken from \cite{Amhis:2019ckw}.
\begin{figure}[tbp!]
    \centering
	\includegraphics[width=0.48\textwidth]{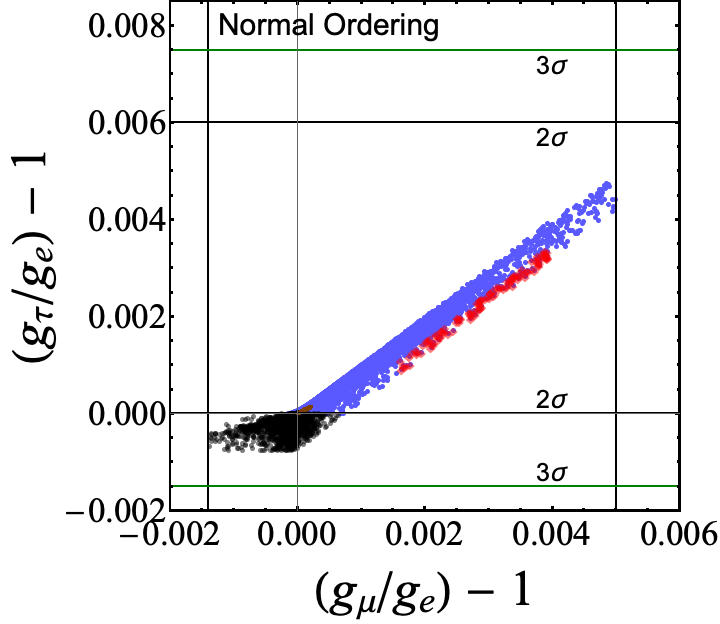}
	\includegraphics[width=0.48\textwidth]{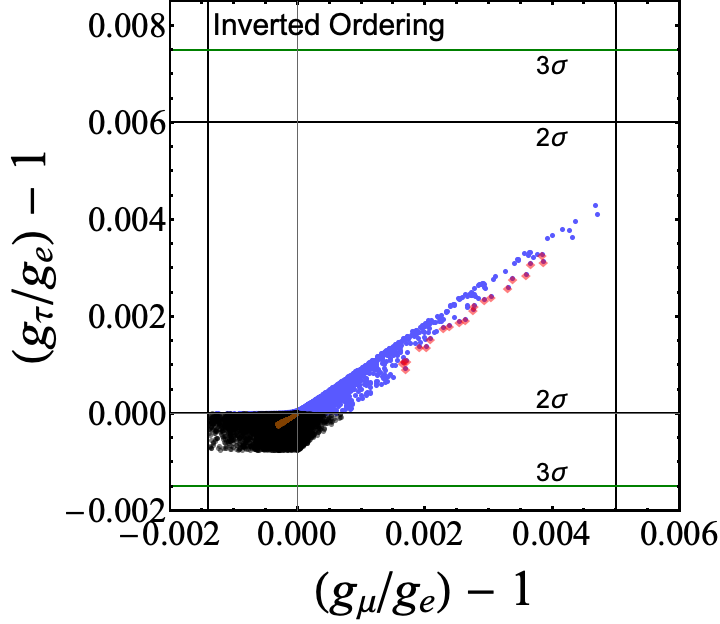}
	\includegraphics[width=0.48\textwidth]{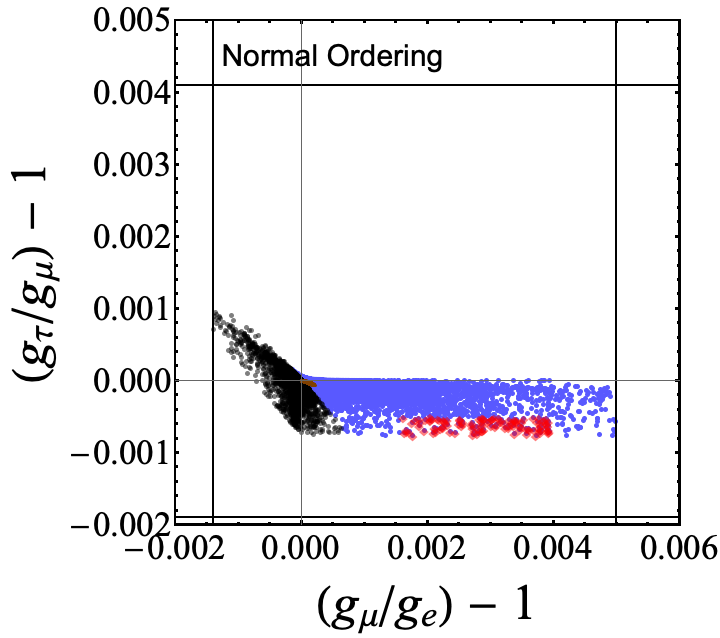}
    \includegraphics[width=0.48\textwidth]{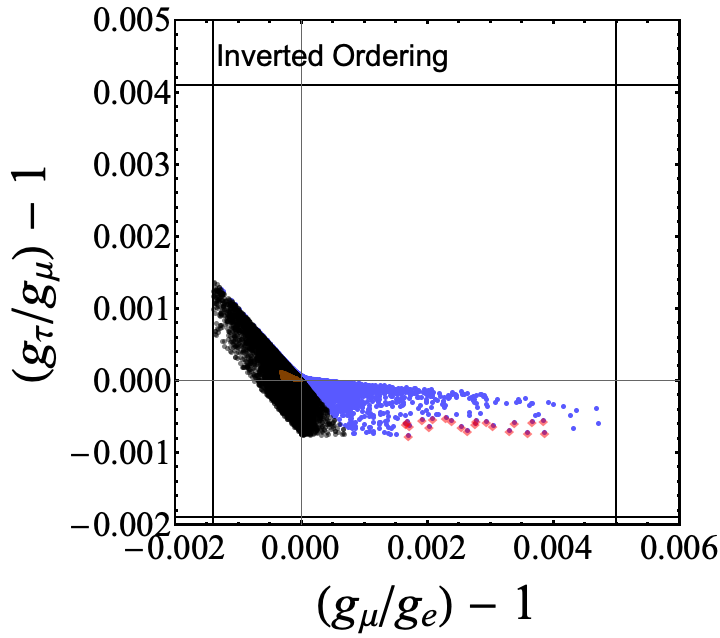}
	\caption{
	Correlations among deviations of $g_\mu/g_e$, $g_\tau/g_e$ (top) and $g_\tau/g_\mu$ (bottom) from universality for NO (left) and IO (right). All shown sample points explain neutrino masses and respect the bounds arising from the flavour observables considered in this work. For the blue points, the deviation of $g_\tau/g_e$ from universality is explained at $2\sigma$ in the linear case (Case I), and for the black points at $3\sigma$, but not at $2\sigma$. The red diamonds also explain the flavour anomalies, for which $g_\tau/g_e$ must be respected at $2\sigma$. Brown points pertain to the quadratic case (Case II) which only occurs at $2\sigma$ for NO and at $3\sigma$ for IO. Solid lines indicate current experimental bounds (black for $2\sigma$ or  $\SI{90}{\percent}$ CL, and green for $3\sigma$).
}
	\label{plot_universality}
\end{figure}

In the top panel of Fig.~\ref{plot_universality} we show $g_\tau/g_e-1$ as a function of $g_\mu/g_e-1$ for NO (left) and IO (right). The results of the numerical scan for the linear case are represented by blue sample points which explain $g_\tau/g_e$ at $2\sigma$, and by black sample points which explain $g_\tau/g_e$ at $3\sigma$, but not at $2\sigma$, see also the caption of Fig.~\ref{plot_universality} for details. The $3\sigma$ region for $g_\tau/g_e$ has been included to accommodate the SM prediction. If not indicated otherwise, ``at $2\sigma$" and ``at $3\sigma$" always refer to this distinction. Red sample points in diamond shape also explain the flavour anomalies which are briefly discussed below. The quadratic case is shown in brown. In Tab.~\ref{tab:observables} we summarise the respective confidence levels at which further experimental bounds are imposed. For the sample points, the same colour code is used throughout this work, except for Figs.~\ref{fig:omno} and \ref{fig:BrvsDelta}.

\emph{We find that there are solutions to the linear-case constraint in Eq.~\eqref{constraint_eigenvector} for both neutrino-mass orderings which simultaneously explain the flavour anomalies introduced above and respect the bounds from the considered flavour observables.} One does in particular not have to assume that $|y^{e\tau}_h|$ is negligibly small.\footnote{In order to avoid the bound from $\mu\to e\gamma$, $y^{e\tau}_h$ was set to zero in \cite{Crivellin:2020klg} which in general is not a viable solution to the constraint in Eq.~(\ref{constraint_eigenvector}) and hence is incompatible with neutrino masses.} \emph{Contrariwise, explaining the flavour anomalies in the quadratic case is not possible.}

In the quadratic case, no large deviations from universality can be generated. In particular, none of the respective $1\sigma$ regions for the $g_a/g_b$ which are currently preferred by experiments can be reached. Still, the corrections to both $g_\mu/g_e - 1$ and $g_\tau/g_e - 1$ are strictly positive (negative) for Normal (Inverted) Ordering in the quadratic case, hence a conclusive experimental determination of one of the signs would rule out one of the mass orderings being generated by $h$. Similarly, positive (negative) corrections to $g_\tau/g_\mu - 1$ are severely disfavoured for NO (IO).

In the linear case,
large contributions to $g_\mu/g_e$ $(g_\tau/g_e)$ are disfavoured for IO as they enforce $|y^{\mu\tau}_h| \gg |y^{e\tau}_h| (|y^{e\mu}_h|)$, see Sect.~\ref{sec:moc} for more details. On the contrary, for IO we find more sample points with $g_\tau/g_\mu>1$ as shown in the bottom panel of Fig.~\ref{plot_universality}. This is due to the fact that a hierarchy between $|y^{e\mu}_h|$ and $|y^{e\tau}_h|$ is easier to achieve in this case. Still, the deviation of $g_\tau/g_\mu$ from universality is measured to be smaller and an explanation of its best-fit value via $h$ would imply a further deviation from the best-fit values of the other two ratios. A given mass $M_h$ fixes the ranges of magnitude of $|y^{\mu\tau}_h|$ and $|y^{e\mu}_h|$ for which the flavour anomalies are explained, as in Eq.~\eqref{eq:CAAeq}. Together with the strict experimental limit on $\text{Br}(\mu\to e\gamma)$ which bounds $|y^{e\tau}_h|$ in terms of $|y^{\mu\tau}_h|$, this determines the relative positions of the red and blue sample points in Fig.~\ref{plot_universality}.

A more precise determination of the lepton-flavour universality ratios $g_a/g_b$ mainly relies on reducing the uncertainties in measurements of the branching ratios $\text{Br}(\tau\to\mu(e)\nu\nu)$ and of the tau lifetime \cite{Pich:2020qna}. An improvement of a factor of ten is suggested in \cite{Abada:2019lih}. Further improvement would rely on determining the tau mass at higher precision, for instance upon running a future tau factory at the production threshold \cite{Pich:2020qna,Abada:2019lih,Barniakov:2019zhx,Luo:2019xqt}. Nonetheless, shifts in the measured values $g_a/g_b$ themselves cannot be predicted and we refrain from showing estimates for prospective sensitivities in Fig.~\ref{plot_universality}.

%%%%%%%%%%%%%%%%%%%%%%%%%%%%%%%%%%%%%%%%%%%%%%%%%%%%%%%%%%%%
\subsubsection*{$W$-Boson Mass}

The contribution to the Fermi constant induced by $h$ results in a necessarily negative correction \cite{Brivio:2017vri}
\begin{align}
    \delta M^2_W = -\frac{M^2_W}{\sqrt{2}G_F}\left|1 - \frac{M_WM_Z}{2M^2_W - M^2_Z}\right|\frac{|y^{e\mu}_h|^2}{M^2_h}
\end{align}
to the $W$-boson mass which exacerbates the existing $\num{1.5}\sigma$ tension among the SM prediction $M_W \pm \Delta M_W = \SI{80.361\pm0.005}{\giga\electronvolt}$ and the world average of measurements given by $M^{\text{exp}}_W \pm \Delta M^{\text{exp}}_W = \SI{80.379\pm0.012}{\giga\electronvolt}$ \cite{Erler:2019ddu,Zyla:2020zbs,Freitas:2020kcn}. In order to accommodate an explanation of the flavour anomalies, we allow for a $3\sigma$ discrepancy which implies $M_W \ge \SI{80.343}{\giga\electronvolt}$ and gives rise to the constraint $|y^{e\mu}_h|^2/M^2_h \lesssim \num{1.25}\times \num{E-2}/\si{\tera\electronvolt\squared}$.
To compare, the best-fit value presented in \cite{Crivellin:2020klg} corresponds to $|y^{e\mu}_h|^2/M^2_h \approx \num{1.07}\times \num{E-2}/\si{\tera\electronvolt\squared}$.
\begin{figure}[btp!]
    \centering
	\includegraphics[width=0.48\textwidth]{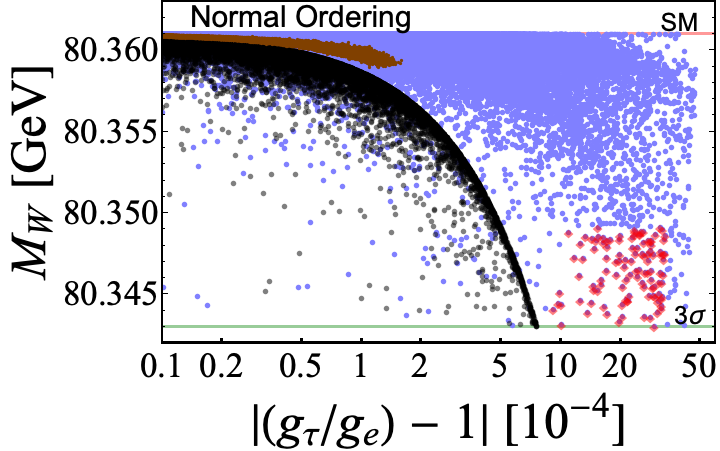}
	\includegraphics[width=0.48\textwidth]{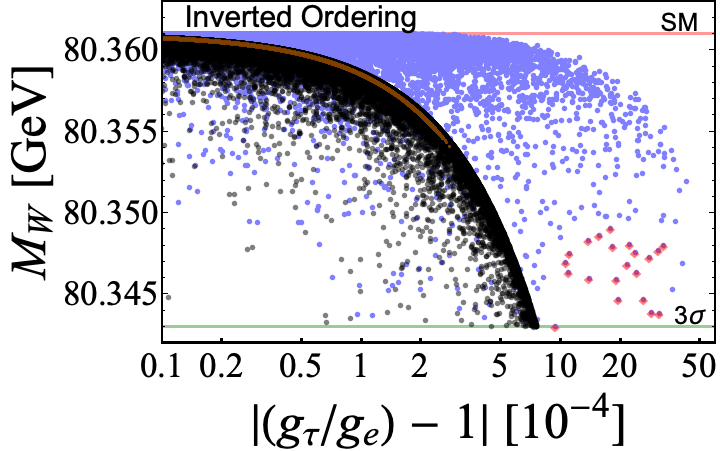}
	\caption{Correlations among $M_W$ and deviations of $g_\tau/g_e$ from universality. The colours are the same as in Fig.~\ref{plot_universality}.}
	\label{fig:mW}
\end{figure}
In Fig.~\ref{fig:mW} the prediction for the $W$-boson mass is shown as a function of the absolute value of the deviation of the universality ratio $g_\tau/g_e$ from one. Note that in the quadratic case the maximum correction to $M_W$ is much larger for IO than for NO, and there is a non-trivial correlation in the linear case especially for $|y^{e\mu}_h| > |y^{\mu\tau}_h|$.
A large effect in $g_\tau/g_e - 1$ together with a conclusive determination of $M_W$ close to its current SM prediction would severely disfavour the scenario of $h$ explaining neutrino masses with IO, but not with NO. Furthermore, a result $M_W \gtrsim \SI{80.35}{\giga\electronvolt}$ would currently rule out an explanation of the flavour anomalies via $h$. In proposals for next-generation lepton colliders, a reduction of the uncertainty in the experimental determination of $M_W$ by a factor of roughly $10 - 20$ \cite{Abada:2019lih,CEPCStudyGroup:2018ghi} is suggested. As for the universality ratios $g_a/g_b$, any shifts in the obtained value $M_W$ itself, be it determined at colliders or via electroweak fits, cannot be predicted though.

\subsubsection*{Leptonic Non-Standard Interactions}
\begin{figure}[b!]
    \centering
	\includegraphics[width=0.48\textwidth]{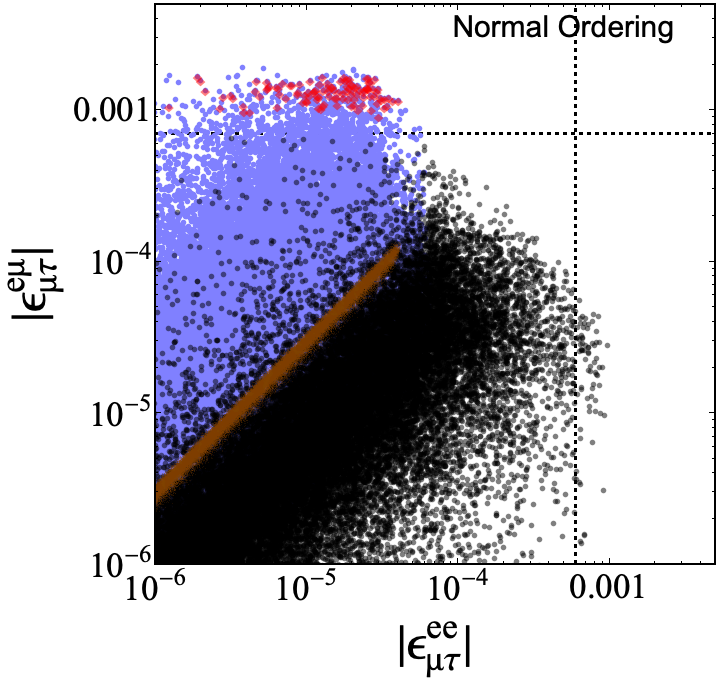}
	\includegraphics[width=0.48\textwidth]{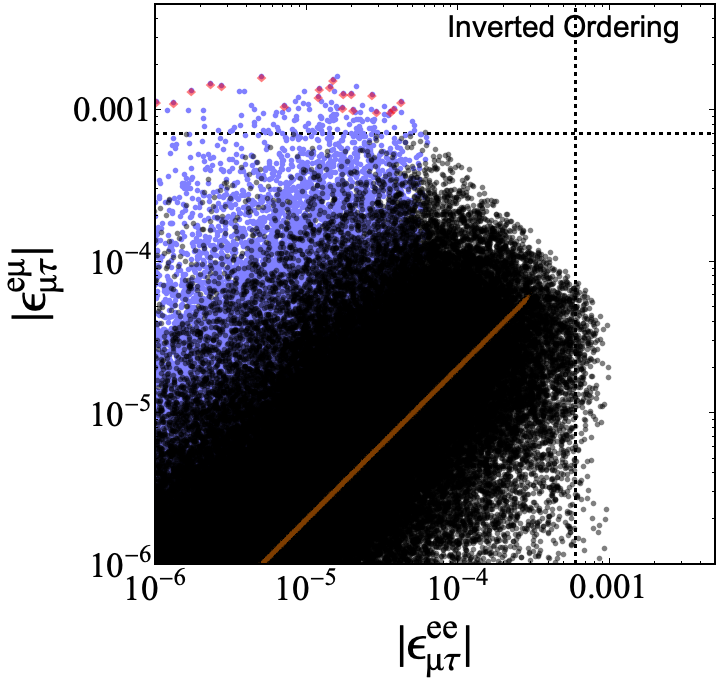}
	\caption{Non-standard interactions. The colours are the same as in Fig.~\ref{plot_universality}. Dashed lines indicate prospected experimental sensitivities.}
	\label{fig:NSI}
\end{figure}
As it can be seen from Eq.~(\ref{eqn:ncl}), the singly-charged scalar singlet induces leptonic non-standard interactions at tree level, whereas NSIs with quarks only arise at loop level. Hence, we disregard the latter. The fact that the constraint in Eq.~\eqref{constraint_eigenvector} disfavours large hierarchies among the coupling magnitudes (see Sect.~\ref{sec:moc} for more details) implies that the results found in studies in which only one NSI parameter was switched on at a time (see for instance \cite{Farzan:2017xzy}) are not directly applicable here. We obtain magnitudes of up to $|\epsilon^{\rho\sigma}_{\alpha\beta}|\sim\num{E-3}$ in the linear case which to our knowledge is below all current bounds and also appears to be challenging to observe in near-future experiments. For  instance, depending on the flavour channel, DUNE is prospected to be sensitive to magnitudes down to $|\epsilon^{\rho\sigma}_{\alpha\beta}|\sim\num{E-2}$ at \SI{90}{\percent} CL \cite{Dev:2018}. Still, at a future neutrino factory it might be possible to probe some of the NSI parameters relevant for neutrino production in the $\nu_e\to\nu_\tau$ and $\nu_\mu\to\nu_\tau$ channels \cite{Herrero-Garcia:2014hfa}:
\begin{align}\label{eqn:nsip}
    \epsilon^{e\mu}_{\tau e} &\equiv \frac{(y^{e\tau}_h)^*y^{e\mu}_h}{\sqrt{2}G_FM^2_h} = -(\epsilon^{ee}_{\mu\tau})^*,
    &\epsilon^{e\mu}_{\mu\tau} &\equiv-\frac{(y^{e\mu}_h)^*y^{\mu\tau}_h}{\sqrt{2}G_FM^2_h}\,.
\end{align}
Upon using a 2 kt OPERA-like near tau detector a sensitivity to $|\epsilon^{e\mu}_{\mu\tau}|\sim\num{7E-4}$ and $|\epsilon^{e\mu}_{\tau e}|\sim\num{6E-4}$ is prospected to be achievable \cite{Ohlsson:2009vk,Tang:2009na}. In that case, the simultaneous explanation of neutrino masses and the flavour anomalies via $h$ in the linear case could be conclusively tested at a neutrino factory for both neutrino-mass orderings. The contributions in the quadratic case will remain beyond reach. This is illustrated in Fig.~\ref{fig:NSI}.
As indicated in Eq.~\eqref{eqn:nsip}, the NSI parameter $\epsilon^{e\mu}_{\tau e}$ is trivially related to the corresponding one for the propagation of $\nu_\mu$ and $\nu_\tau$ neutrinos in matter.

\subsection{Charged Lepton Flavour Violation}

\begin{figure}[hbpt!]
    \centering
	\includegraphics[width=0.48\textwidth]{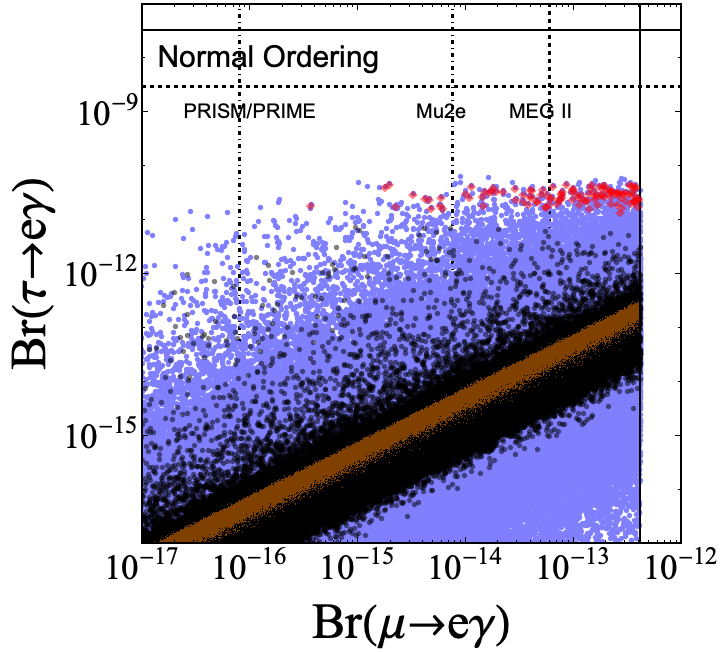}
	\includegraphics[width=0.48\textwidth]{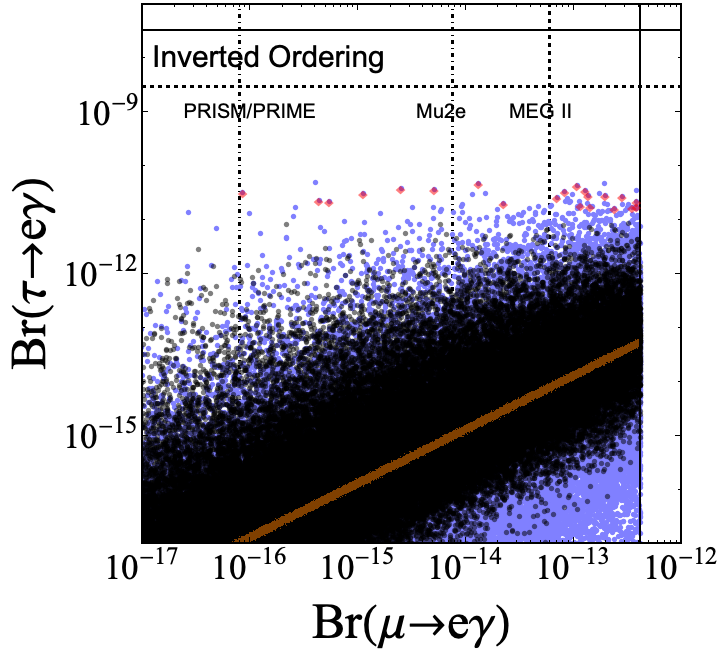}
	\includegraphics[width=0.48\textwidth]{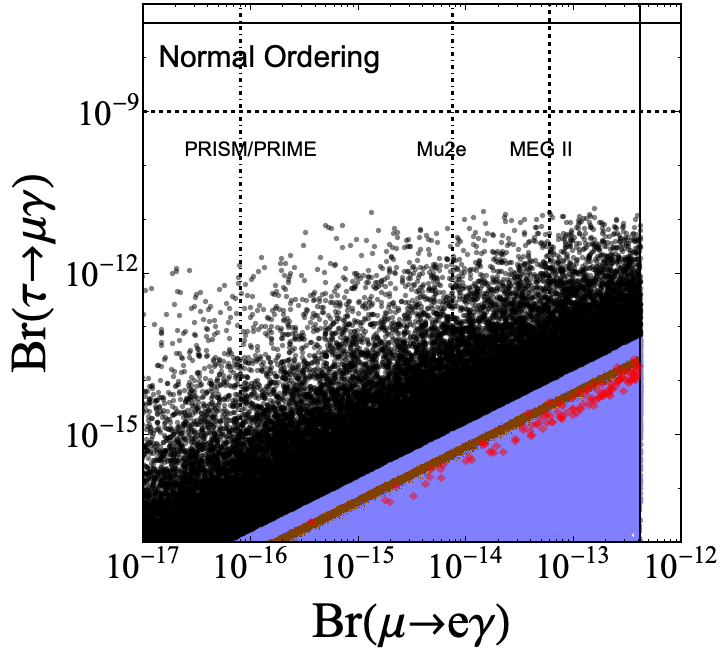}
	\includegraphics[width=0.48\textwidth]{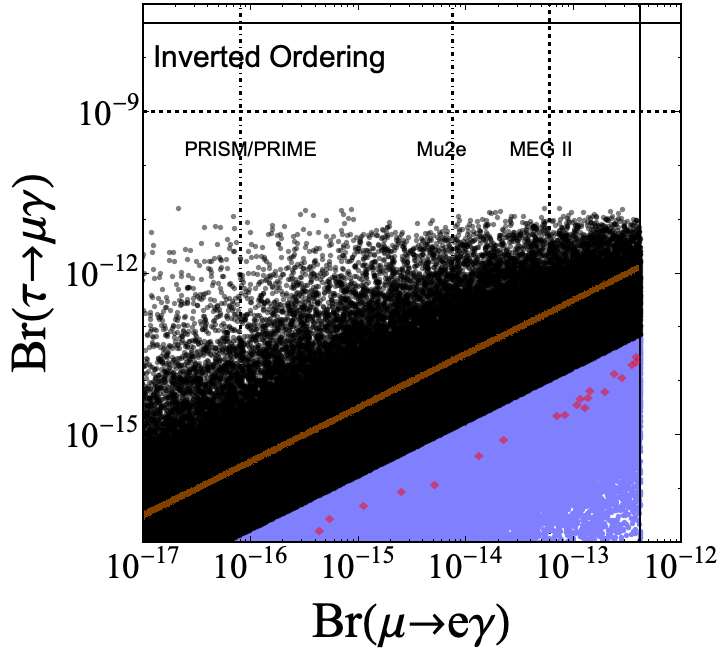}
	\caption{Branching ratios of radiative charged-lepton decays. The vertical dot-dashed lines correspond to the prospected sensitivities to $\text{Br}(\mu\to e\text{;}\;\text{Ti})$ at PRISM/PRIME and to $\text{Br}(\mu\to e\text{;}\;\text{Al})$ at Mu2e which were then converted via $\text{Br}(\mu\to e\text{;}\;\text{Al})\approx\num{0.0079}\,\text{Br}(\mu\to e\gamma)$ and $\text{Br}(\mu\to e\text{;}\;\text{Ti})\approx\num{0.0125}\,\text{Br}(\mu\to e\gamma)$, see also App.~\ref{sec:Mu2E}. The colours are the same as in Fig.~\ref{plot_universality}. \label{fig:llgamma}}
\end{figure}

The leading-order contributions to flavour-violating charged-lepton decays from singly-charged scalar singlets occur at one-loop level. In fact, finite contributions to radiative charged-lepton decays $\ell_i\to \ell_j\gamma$ are sourced by a single diagram with a neutrino $\nu_k$, $i\neq j$ and $i\neq k\neq j$,
in the loop. The branching ratios are given by~\cite{PETCOV1982401,Nebot:2007bc,BERTOLINI1988714,RevModPhys.59.671,PICH1984229,Crivellin:2020klg}
\begin{align}
    \text{Br}(\mu\to e\gamma) & = \text{Br}(\mu\to e\nu\bar\nu)\frac{\alpha_{\text{EM}}}{48\pi G^2_F}\frac{|y^{e\tau}_hy^{\mu\tau}_h|^2}{M^4_h}, \\
    \text{Br}(\tau\to e\gamma) & = \text{Br}(\tau\to e\nu\bar\nu)\frac{\alpha_{\text{EM}}}{48\pi G^2_F}\frac{|y^{e\mu}_hy^{\mu\tau}_h|^2}{M^4_h}, \\
    \text{Br}(\tau\to \mu\gamma) & =  \text{Br}(\tau\to \mu\nu\bar\nu)\frac{\alpha_{\text{EM}}}{48\pi G^2_F}\frac{|y^{e\mu}_hy^{e\tau}_h|^2}{M^4_h}\,,
\end{align}
with $\text{Br}(\mu\to e\nu\bar\nu)\approx 1$, $\text{Br}(\tau\to e\nu\bar\nu)\approx\num{0.178}$ and $\text{Br}(\tau\to\mu\nu\bar\nu)\approx\num{0.174}$ \cite{Zyla:2020zbs}. As it can be seen in Fig.~\ref{fig:llgamma}, any signal in radiative tau decays showing up at Belle II cannot be induced by $h$ alone, see Table~\ref{tab:observables}. For instance, one would need two singly-charged scalar singlets which conspire to circumvent the strong bounds arising from flavour-violating muon decays.
Also the sizeable contributions to $\tau\to e\gamma$ implied by simultaneously generating neutrino mass and explaining the flavour anomalies will be beyond reach \cite{Crivellin:2020klg}. Instead, a future search for $\mu\to e\gamma$ \cite{Baldini:2018nnn} efficiently probes parts of the parameter space pertaining to $h$ generating neutrino masses both in the linear case and in the quadratic case, as well as the combined scenario in which also the flavour anomalies are explained in the linear case.

\begin{figure}[tb!]
    \centering
	\includegraphics[width=0.48\textwidth]{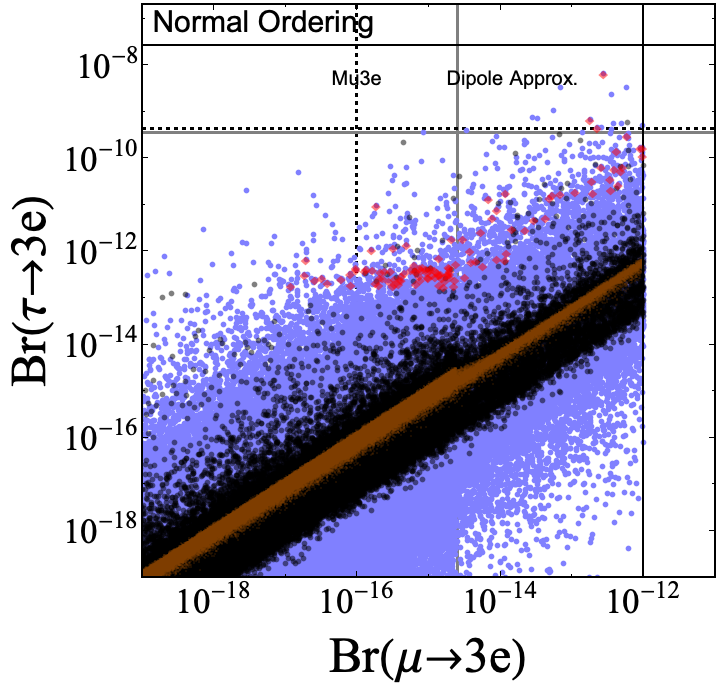}
	\includegraphics[width=0.48\textwidth]{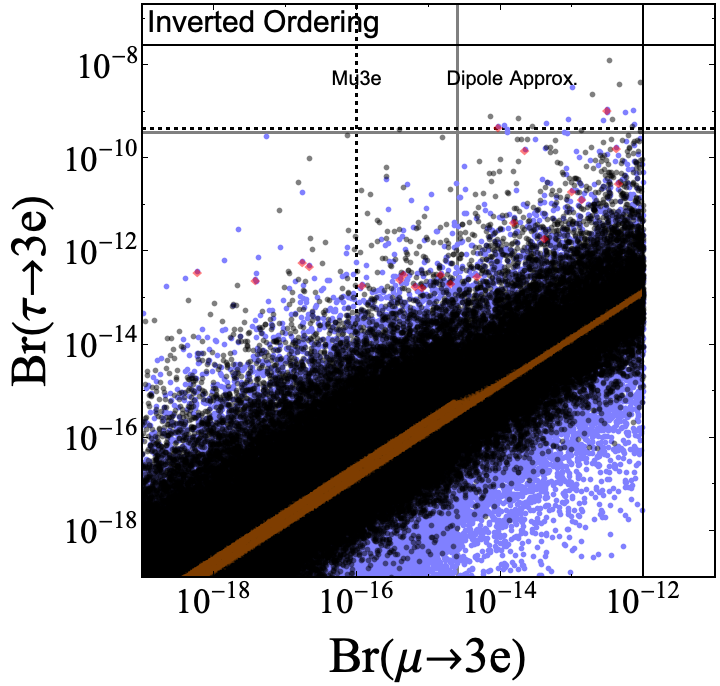}
	\includegraphics[width=0.48\textwidth]{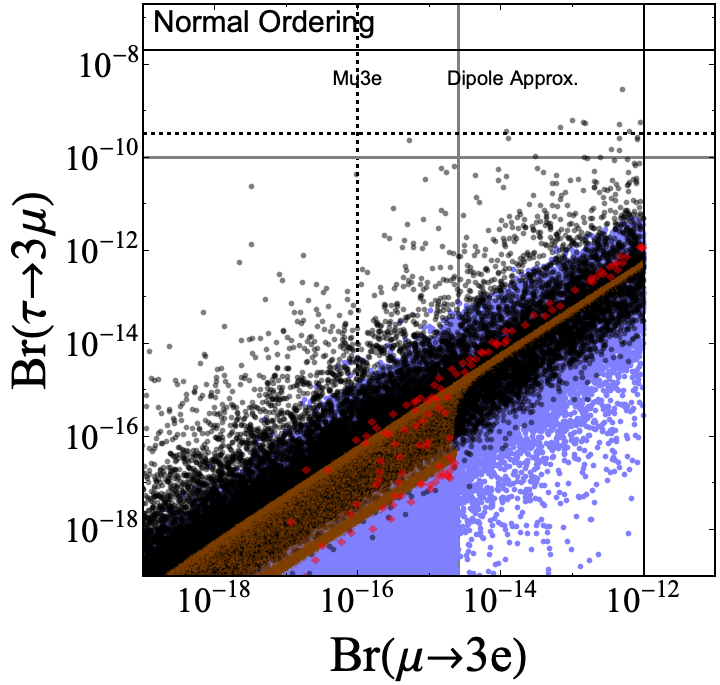}
	\includegraphics[width=0.48\textwidth]{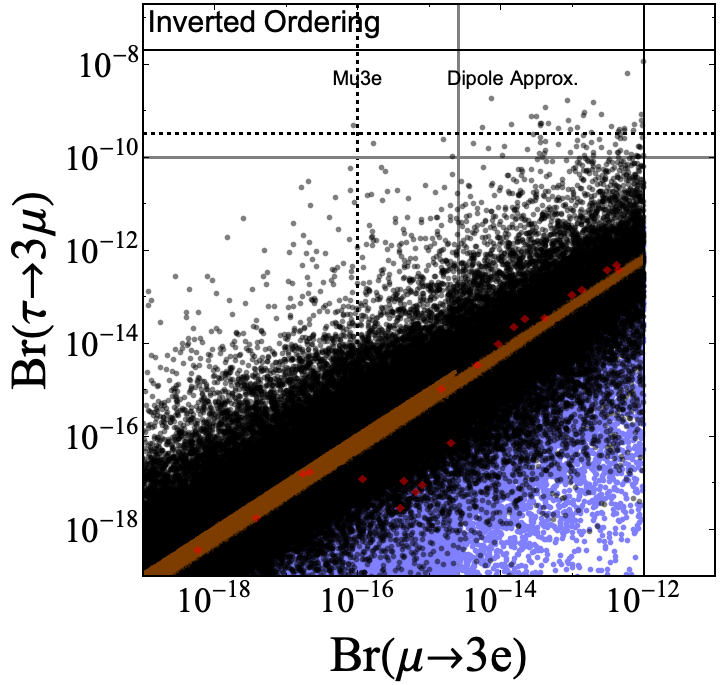}
	\caption{Branching ratios of tri-lepton decays. The colours are the same as in Fig.~\ref{plot_universality}. The solid grey lines indicate the respective experimental bounds that would apply to the photon-penguin approximation. Dashed lines indicate prospected experimental sensitivities.}
	\label{plot_trilepton_decays}
\end{figure}

As it is well-known, if the contributions from on-shell photon penguin diagrams dominate, the branching ratios for tri-lepton decays with only one flavour in the final state are entirely fixed as functions of $\text{Br}(\ell_i\to \ell_j\gamma)$ and SM parameters:\footnote{We do not expect more stringent constraints from tri-lepton decays with different flavours in the final state and hence we do not consider them.}
\cite{Kuno:1999jp,Crivellin:2013hpa}
\begin{align}
    \frac{\text{Br}(\mu\to3e)}{\text{Br}(\mu\to e\gamma)} & \approx \frac{\alpha_{\text{EM}}}{3\pi}\left(\log\left(\frac{m^2_\mu}{m^2_e}\right) - \frac{11}{4}\right) \approx \frac{1}{163}, \\
    \frac{\text{Br}(\tau\to3e)}{\text{Br}(\tau\to e\gamma)} & \approx \frac{\alpha_{\text{EM}}}{3\pi}\left(\log\left(\frac{m^2_\tau}{m^2_e}\right) - \frac{11}{4}\right) \approx \frac{1}{95}, \\
    \frac{\text{Br}(\tau\to3\mu)}{\text{Br}(\tau\to \mu\gamma)} & \approx \frac{\alpha_{\text{EM}}}{3\pi}\left(\log\left(\frac{m^2_\tau}{m^2_\mu}\right) - \frac{11}{4}\right) \approx \frac{1}{446}\,.
\end{align}
For masses close to the lower bound $M_h = \SI{350}{\giga\electronvolt}$, the photon-penguin approximation is perfectly valid.
In the quadratic case, the relative magnitudes of the couplings $y^{ij}_h$ are quite sensitive to the neutrino-mass ordering, as dictated in Eq.~\eqref{constraint_eigenvector_special}. Together with the flavour-dependent suppression factors $\sim\log(m_k/m_l)$, this efficiently determines the relative size of the different radiative charged-lepton decay channels in the photon-penguin limit. On the contrary, note how the contributions from box diagrams outperform those from photon penguins for $\tau\to3\mu$ in the case of NO, as can be seen in Fig.~\ref{plot_trilepton_decays}.

\begin{figure}[hbpt!]
    \centering
	\includegraphics[width=0.48\textwidth]{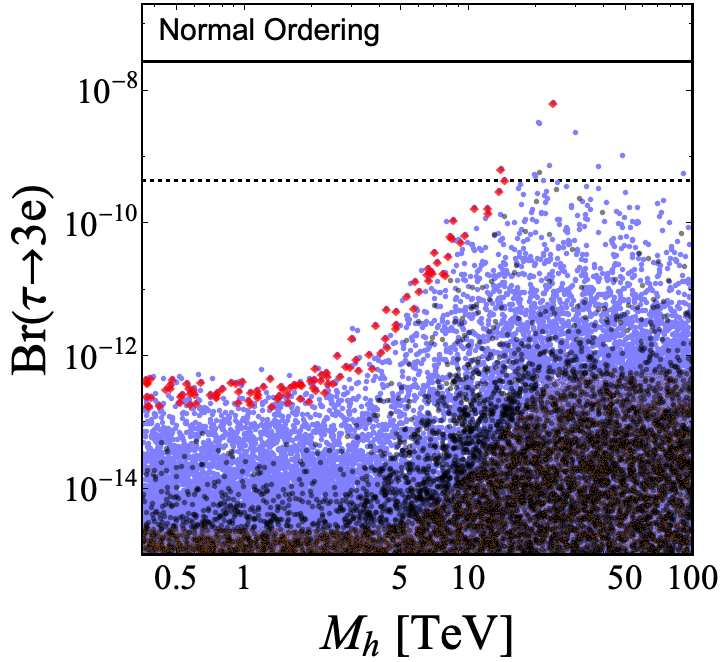}
	\includegraphics[width=0.48\textwidth]{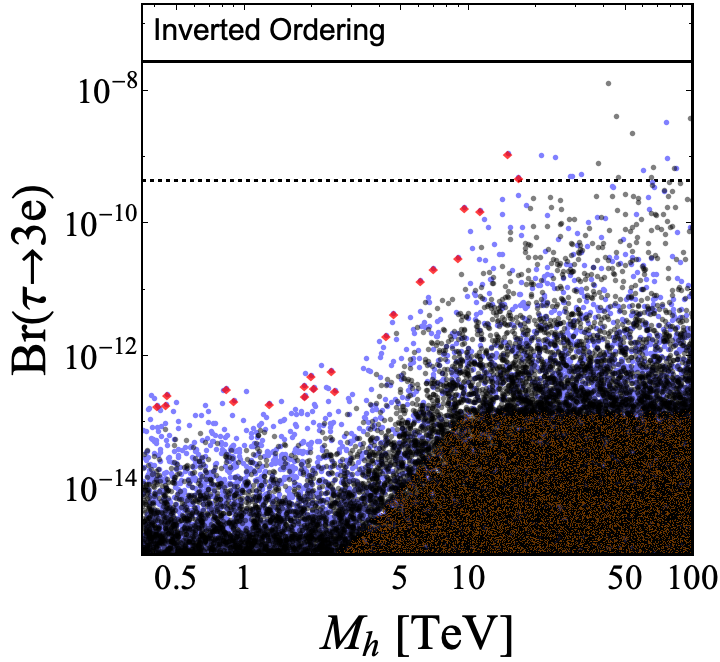}
	\includegraphics[width=0.48\textwidth]{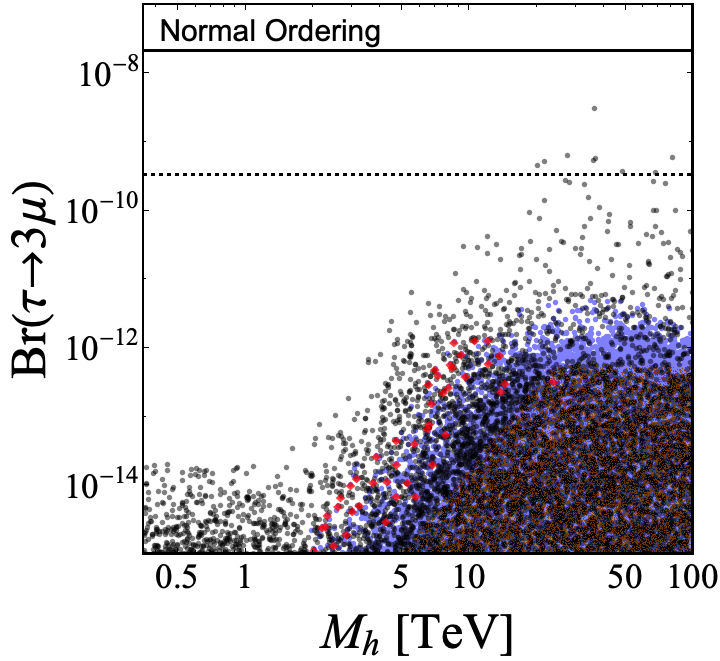}
	\includegraphics[width=0.48\textwidth]{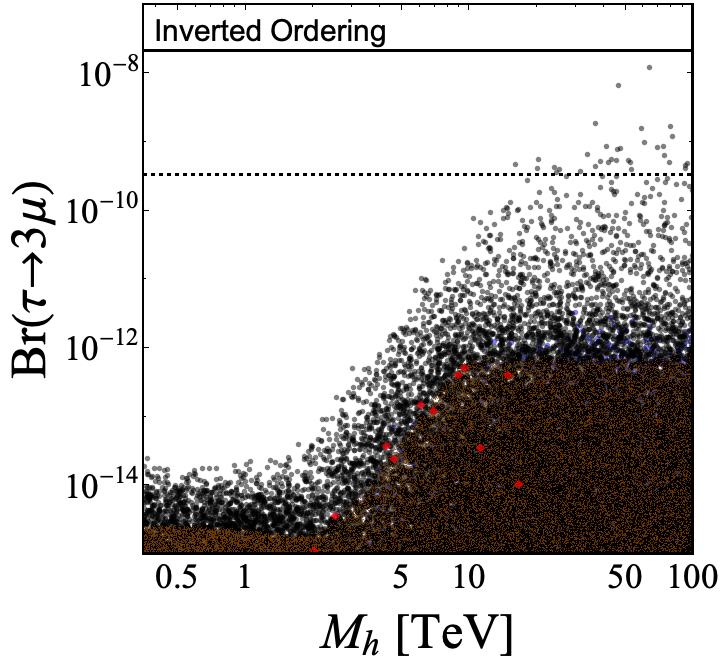}
	\caption{Branching ratio of tri-lepton tau decays as a function of the mass $M_h$. The colours are the same as in Fig.~\ref{plot_universality}. Dashed lines indicate prospected experimental sensitivities.}
	\label{fig:tau2eee}
\end{figure}

The vertical solid grey lines in Fig.~\ref{plot_trilepton_decays} indicate the bound induced by $\mu\to e\gamma$ which is the relevant one both for the linear case and for the quadratic case as long as the photon penguin dominates $\mu\to 3e$.
In the numerical scan, the full expression as given in \cite{Crivellin:2020klg} is used, because larger masses $M_h$ generally render larger magnitudes $|y^{ij}_h|$ compatible with the different experimental bounds, which in turn implies that the contributions from box diagrams to tri-lepton decays become increasingly dominant. Since box diagrams are proportional to the product of four Yukawa couplings, they can thus induce contributions to tri-lepton decays which in fact grow if the mass increases beyond $M_h \approx\SI{1}{\tera\electronvolt}$ and further. Thus, $h$ will decouple from the phenomenology at low energy only for even larger masses $M_h\gg\SI{100}{\tera\electronvolt}$. This is distinctively visible in Fig.~\ref{fig:tau2eee} where we show $\tau\to 3e(\mu)$ as a function of the singly-charged scalar singlet mass $M_h$.
As the figures illustrate, masses larger than $M_h\gtrsim\SI{11}{\tera\electronvolt}$ and $M_h\gtrsim\SI{15}{\tera\electronvolt}$ can induce an observable signal in $\tau\to 3e$ and $\tau\to 3\mu$, respectively, at Belle II \cite{Kou:2018nap}.

Besides $\mu\to 3e$ \cite{Arndt:2020obb} which will be sensitive both to the linear case and the quadratic case, tri-lepton tau decay thus offer another avenue for testing the generation of neutrino masses via $h$ at larger masses $M_h$ in the linear case, complementary to $\mu\to e\gamma$ for which the contributions start to decrease before the assumed upper limit $M_h \le\SI{100}{\tera\electronvolt}$ is reached. Via $\tau\to 3e$ we are even sensitive to parts of the parameter space for which the flavour anomalies are explained as well \cite{Crivellin:2020klg}. Still, the constraint in Eq.~\eqref{constraint_eigenvector} disfavours the solutions $y^{ij}_h$ which induce large contributions to $\tau\to 3e(\mu)$ as there needs to be a hierarchy among the coupling magnitudes $|y^{e\mu}_h|$ and $|y^{\mu\tau}_h|$ ($|y^{e\tau}_h|$) entering the relevant box diagram and $|y^{e\tau}_h|$ ($|y^{\mu\tau}_h|$) which must be smaller in order not to violate the experimental bound on $\text{Br}(\mu\to e\gamma)$.

A further relevant process is $\mu - e$ conversion in nuclei which probes the same parameter combination as $\mu\to e\gamma$ and is dominated by photon-penguin diagrams. As of today, the strongest constraint arises from the SINDRUM-II experiment in which a gold target was used \cite{Bertl:2006up}. Taking into account both the short-range and the long-range contribution (see App.~\ref{sec:Mu2E}), one finds $\text{Br}(\mu\to e\text{;}\;\text{Au}) \equiv \omega^\text{Au}_{\text{conv}}/\omega^\text{Au}_{\text{capt}} \approx\num{0.0130}\;\text{Br}(\mu\to e\gamma)$ for $\mu-e$ conversion in gold \cite{Crivellin:2018qmi,Kitano:2002mt,Crivellin:2020klg}. Hence, the process does not yield a competitive constraint yet, still, in the photon-penguin approximation it is less suppressed with respect to $\mu\to e\gamma$ than $\mu\to 3e$. In addition, future experiments on $\mu-e$ conversion are prospected to outperform current and future searches for radiative charged-lepton decays in sensitivity by far \cite{Kutschke:2011ux,Hungerford:2009zz,Kuno:2005mm}. For instance, PRISM/PRIME can be expected to almost conclusively test the simultaneous explanation of neutrino masses and the flavour anomalies.

As a side note, the singly-charged scalar singlet also
generates contributions to anomalous magnetic moments.
However, the contribution is always negative
\cite{Nebot:2007bc,1985PhRvD..31..105M,LEVEILLE197863,Lindner:2016bgg},
hence it is not possible to explain the long-standing anomaly
$\delta a_\mu \equiv a^{\text{exp}}_\mu - a^{\text{SM}}_\mu
=(2.51\pm0.59)\times 10^{-9}$
\cite{Bennett:2006fi,Aoyama:2020ynm,Abi:2021gix}.\footnote{See e.g. \cite{Bigaran:2020jil} for an explanation in a new-physics scenario.} We find contributions to the anomalous magnetic moment of the muon of up to $\delta a_\mu \approx \num{-E-11}$. Similarly, in the electron case, the contribution $\delta a_e\approx\num{-E-16}$ is small compared to the experimental uncertainty which is of order $\num{E-13}$~\cite{Hanneke:2008tm,Hanneke:2010au,Parker:2018vye,Morel:2020dww}.

%%%%%%%%%%%%%%%%%%%%%%%%%%%%%%%%%%%%%%%%%%%%%%%%%%%%%%%%%%%%

\subsection{Magnitude of Couplings}
\label{sec:moc}

\begin{figure}[tb!]
    \centering
	\includegraphics[width=0.49\textwidth]{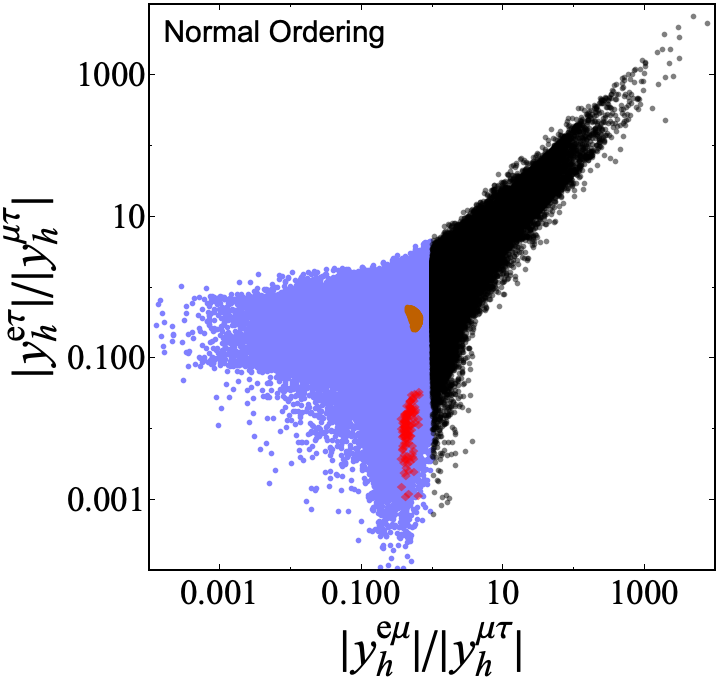}
	\includegraphics[width=0.49\textwidth]{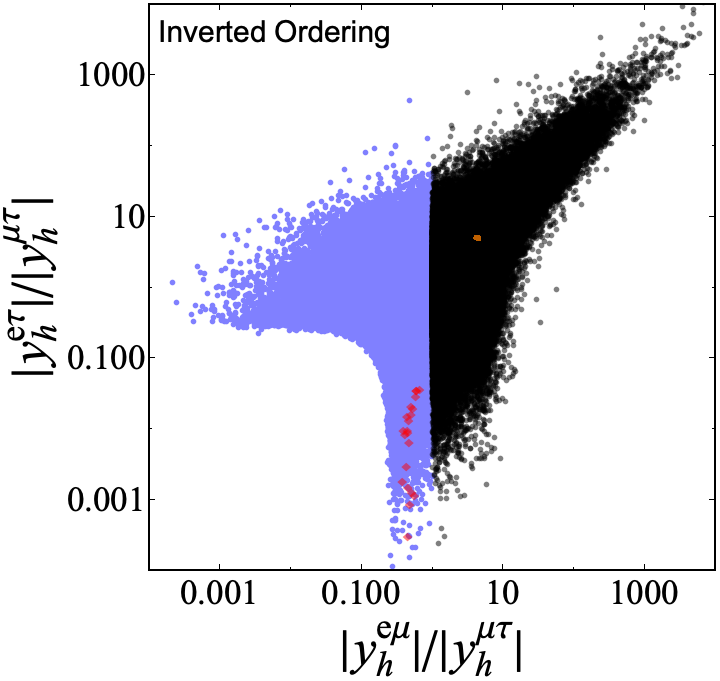}
	\caption{Plot of the coupling ratios $|y^{e\mu}_h|/|y^{\mu\tau}_h|$ and $|y^{e\tau}_h|/|y^{\mu\tau}_h|$ as obtained in the numerical scan. The colours are the same as in Fig.~\ref{plot_universality}.}.
	\label{plot_ratios_couplings}
\end{figure}

The constraint in Eq.~\eqref{constraint_eigenvector} tends to correlate the couplings $y^{ij}_h$ in such a way that in many cases at least two of them are comparable in magnitude, as it can be seen in Fig.~\ref{plot_ratios_couplings}.
We show the ratios because the magnitudes $|y^{ij}_h|$ of the elements of $v_h$ in Eq.~(\ref{constraint_eigenvector}) can always be rescaled by a common factor and hence only the relative magnitudes are determined via the constraint. It is distinctively visible how viable parameter space opens up upon replacing the condition in Eq.~(\ref{constraint_eigenvector_special}) by the more general one in Eq.~(\ref{constraint_eigenvector}).

For NO, solutions with $|y^{\mu\tau}_h|$ larger than both $|y^{ei}_h|$, $i=\mu,\tau$ are most abundant and in particular the hierarchies $|y^{e\mu}_h| < |y^{\mu\tau}_h| < |y^{e\tau}_h|$ and $|y^{e\tau}_h| < |y^{\mu\tau}_h| < |y^{e\mu}_h|$ are rather disfavoured, hence, there is a tendency for $|y^{\mu\tau}_h| \gtrsim |y^{e\mu}_h| \approx |y^{e\tau}_h|$. On the contrary, for IO there are smaller differences in how often the different hierarchies are obtained. Note that while the viable regions in parameter space in Fig.~\ref{plot_ratios_couplings} do in general not feature a sharp contour, the most distinctive deviation from that tendency occurs if both $|y^{e\mu}_h|/|y^{\mu\tau}_h|\lesssim\num{0.1}$ and $|y^{e\tau}_h|/|y^{\mu\tau}_h|\lesssim\num{0.1}$ for which viable solutions seem to be rigorously excluded in the case of IO. Hence, if the coupling $y^{\mu\tau}_h$ was experimentally confirmed to sufficiently dominate over the electron-flavoured ones in magnitude, this would appear to leave us only with the possibility of $h$ generating the main contribution to neutrino masses with NO. The corresponding experimental signature would be a vanishingly small branching ratio for the decay channel $h\to e\nu$, see also Sect.~\ref{sec:dcss}.
\begin{figure}[bt]
    \centering
	\includegraphics[width=0.49\textwidth]{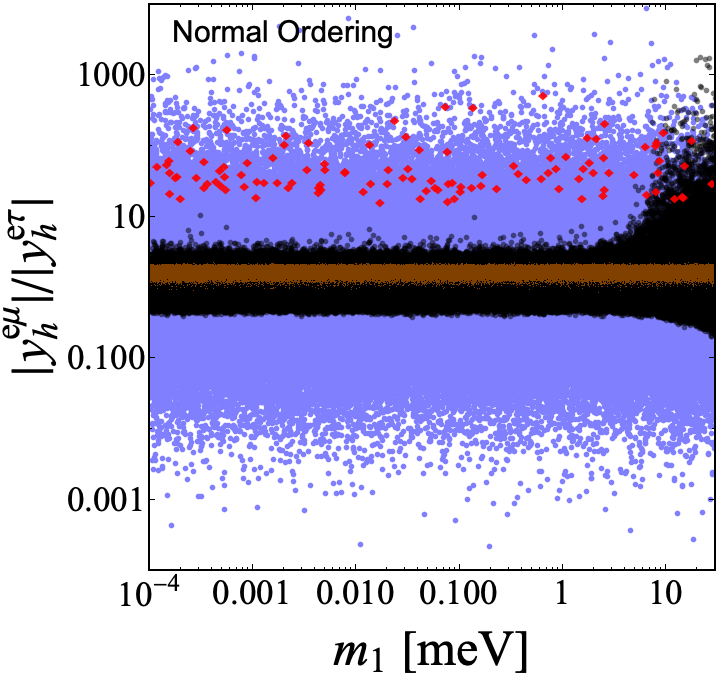}
	\includegraphics[width=0.49\textwidth]{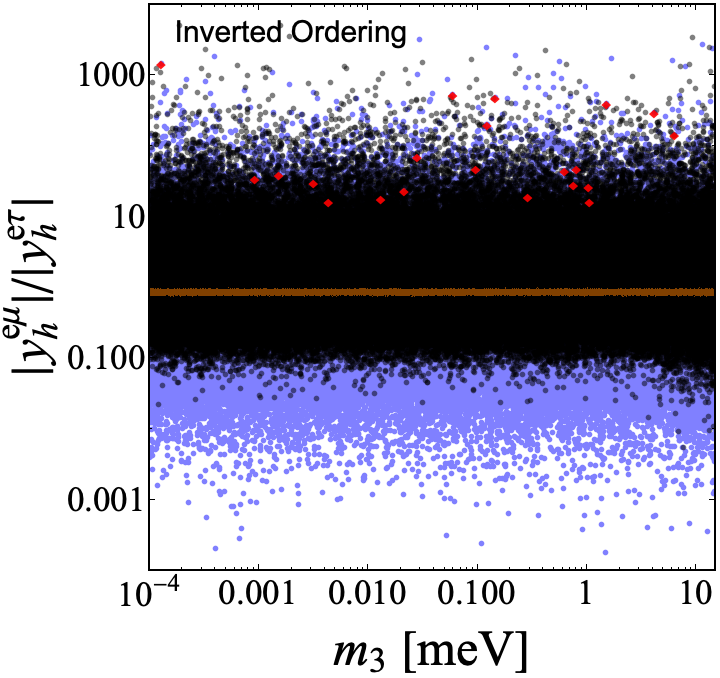}
	\caption{Plot of the coupling ratios $|y^{e\mu}_h|/|y^{e\tau}_h|$ as a function of the smallest neutrino mass ($m_1$ for NO, and $m_3$ for IO). The colours are the same as in Fig.~\ref{plot_universality}.}
	\label{plot_couplings_neutrino_mass}
\end{figure}

Besides, as it can be seen in Fig.~\ref{plot_couplings_neutrino_mass}, if $|y^{e\mu}_h| > |y^{\mu\tau}_h|$, which corresponds to black sample points, the constraint in Eq.~\eqref{constraint_eigenvector} further disfavours solutions with $|y^{e\mu}_h|/|y^{e\tau}_h|\lesssim\num{0.1}$. In addition, if the lightest neutrino is not much heavier than $m_0 = m_1 \approx \SI{1}{\milli\electronvolt}$ in the case of NO, $|y^{e\mu}_h| > |y^{\mu\tau}_h|$ is only viable for $|y^{e\mu}_h|/|y^{e\tau}_h|\lesssim\num{10}$. We trace this back to the fact that for NO the neutrino mass matrix is known to feature a slight hierarchy between the magnitudes of the components in the first row (and column) and those in the 23-block, which only diminishes if the smallest neutrino mass $m_1$ becomes large.

Of course, these solutions are not obtained if one solves the constraint in Eq.~\eqref{constraint_eigenvector} with the additional condition of one Majorana phase and the smallest neutrino mass vanishing. Still, there are no major differences in the obtained phenomenology compared to the general case with three massive neutrinos. In particular, one does not enjoy the same predictive power as in the quadratic case for which the smallest neutrino mass vanishes, $m_0 = 0$, automatically. On the contrary, for IO the $|M^{ij}_\nu|$ are more similar in magnitude and less sensitive to $m_3$, and thus so are the $|y^{ij}_h|$.
\begin{figure}[tb!]
    \centering
	\includegraphics[width=0.5\textwidth]{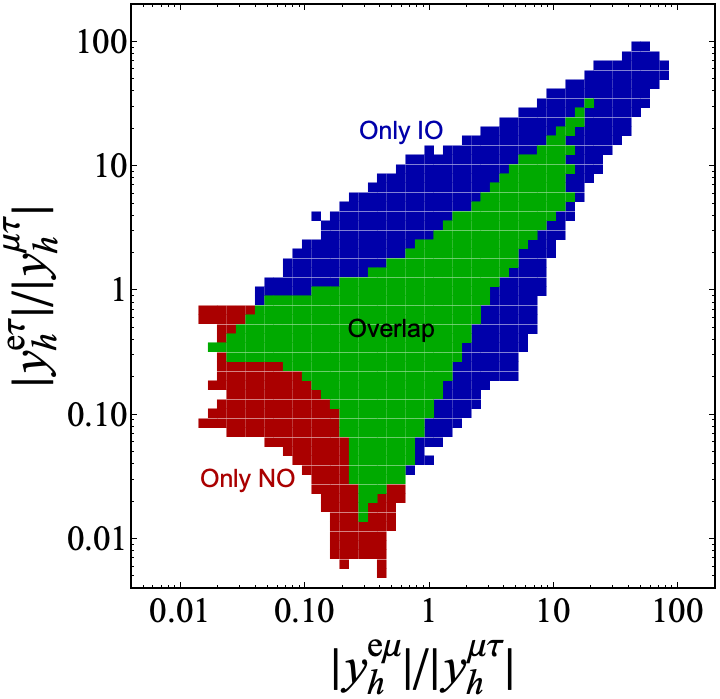}
	\caption{Plot of the coupling ratios $|y^{e\mu}_h|/|y^{\mu\tau}_h|$ and $|y^{e\tau}_h|/|y^{\mu\tau}_h|$ as obtained in the numerical scan if approximately \SI{95.45}{\percent} of the overall number of $547991$ ($542287$) sample points generated for NO (IO) are taken into account. Each square shown to be compatible with NO (IO) contains at least 97 (74) sample points. See also main text.}
	\label{fig:omno}
\end{figure}

Furthermore, a determination of the relative size of the regions in the parameter space of coupling magnitudes which are compatible only with NO or with IO, or with both is performed. The strategy is to discretise the parameter space into a grid structure and to count the sample points contained in each grid square, starting with the square containing the largest number of points and then gradually moving on to those with fewer points, until a specified portion of the overall number of sample points is taken into account.
Fig.~\ref{fig:omno} shows the region of approximately \SI{95.45}{\percent} of the sample points.

\subsection{Decay Channels of the Singly-Charged Scalar Singlet}
\label{sec:dcss}

The partial width of the decay of a singly-charged scalar singlet into a charged lepton $\ell$ and a neutrino $\nu$ is given by \cite{Nebot:2007bc}
\begin{align}
    \Gamma(h\to \ell_a\nu_b) = \Gamma(h\to \ell_b\nu_a) = \frac{|y^{ab}_h|^2}{4\pi}M_h\,.
\end{align}
Leaving the undetected neutrino flavour unspecified, one obtains the branching ratio for the decay of $h$ into a charged lepton of flavour $a$ and a neutrino:
\begin{align}
    \text{Br}(h\to \ell_a\nu) = \frac{\sum_{b\neq a} |y^{ab}_h|^2}{2(|y_h^{e\mu}|^2+|y_h^{e\tau}|^2+|y_h^{\mu\tau}|^2)}\,.
\end{align}
Regardless of whether the magnitudes $|y^{ij}_h|$ are constrained in some way or not, the individual branching ratios always take a value between 0 and \num{0.5}.
\begin{figure}[tb]
    \centering
    \includegraphics[width=0.495\textwidth]{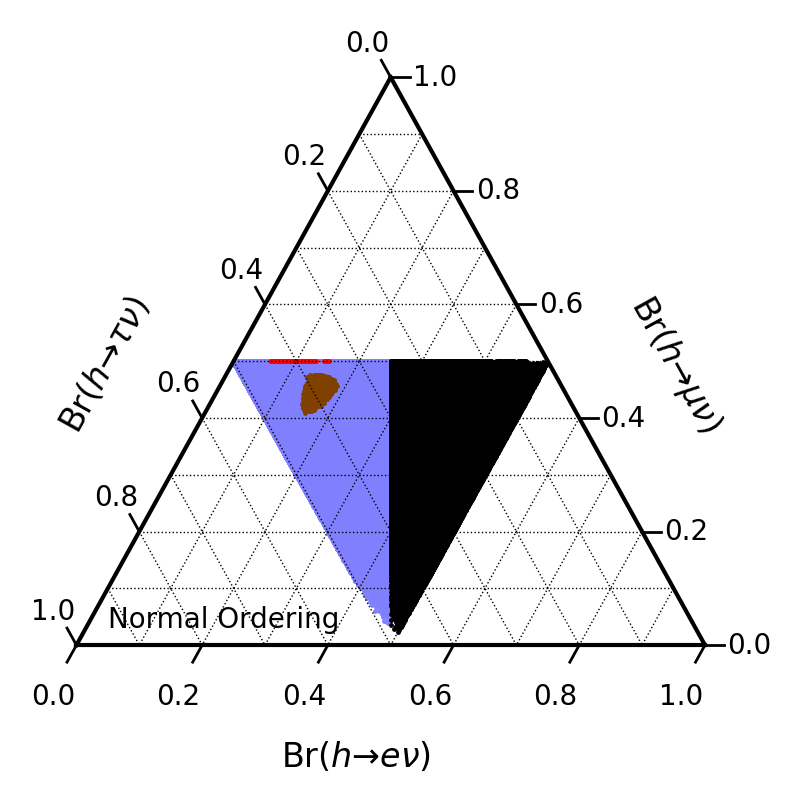}%
	\includegraphics[width=0.495\textwidth]{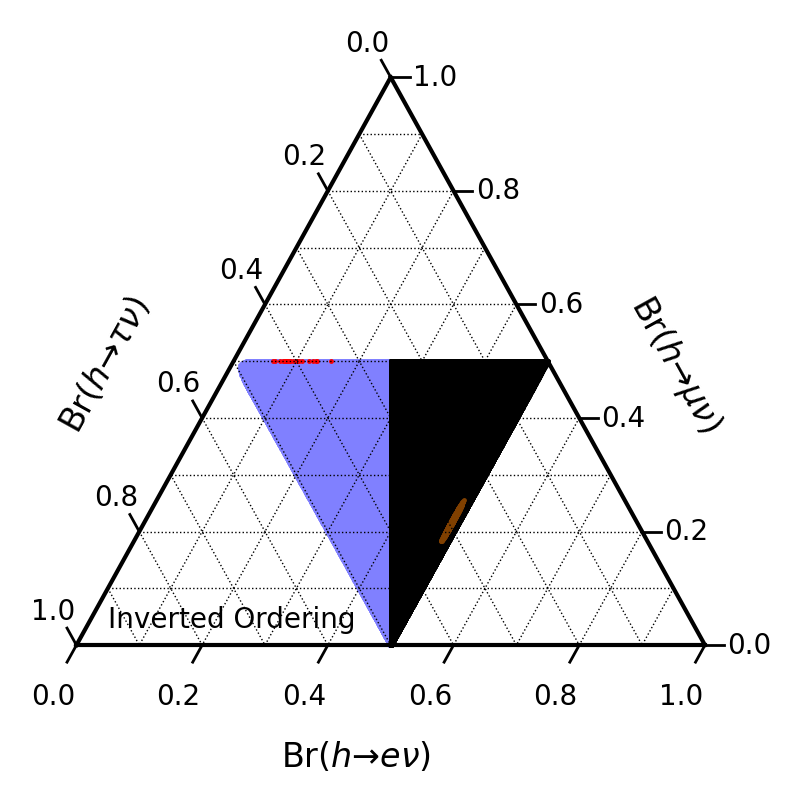}
\caption{Branching ratios of $h\to \ell \nu$. The colours are the same as in Fig.~\ref{plot_universality}.}
	\label{fig:Brlnu}
\end{figure}%
\begin{figure}[tb]
    \centering
	\includegraphics[width=0.7\textwidth]{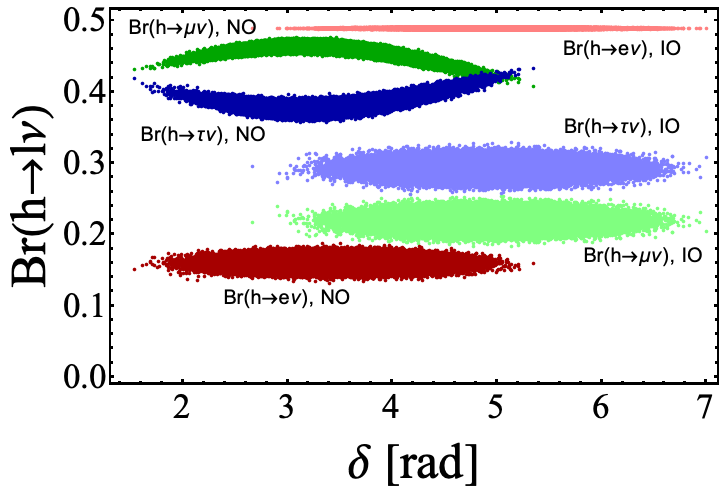}
	\caption{Branching ratios of the different channels $h\to \ell \nu$ as functions of the Dirac CP phase $\delta$ in the quadratic case (Case II).}
	\label{fig:BrvsDelta}
\end{figure}

A very small or vanishing branching ratio for the electron channel (and consequently $\text{Br}(h\to\mu\nu) \approx \text{Br}(h\to\tau\nu) \approx \num{0.5}$) is supported in NO, but severely disfavoured for IO, as it can be seen in Fig.~\ref{fig:Brlnu}. Contrariwise, obtaining a (near-)maximal branching to taus and a small one to muons is disfavoured by NO, but compatible with IO. For the more restrictive constraint in Eq.~(\ref{constraint_eigenvector_special}), the muon and tau channels exhibit a slight correlation with the CP-violating phase $\delta$ in the case of NO which is illustrated in Fig.~\ref{fig:BrvsDelta} and has been discussed before in \cite{Herrero-Garcia:2014hfa} for the Zee-Babu model. It is due to the fact that for IO $\delta$ only fixes the phases $\arg(y^{ei}_h)$, while it also determines $|y^{\mu\tau}_h|$ in the case of NO.

\section{Multiple Singly-Charged Scalar Singlets}
\label{sec:more_singly_charged}

Lastly, we comment on the possibility of generating $1,...,n$ sizeable contributions to neutrino masses via multiple singly-charged scalar singlets $h_1,...,h_n$ and focus on $n = 2$ for simplicity.

The overall lepton sector for two singly-charged singlet scalars is given by
\begin{align}\label{leptonic_Lagrangian_CaseIII}
    \mathcal{L}_{\text{lept}} = y^{ij}_e\bar e_iL_jH^* + y^{ij}_{h_1}L_iL_jh_1 + y^{ij}_{h_2}L_iL_jh_2 + \text{h.c.}.
\end{align}
We again can distinguish the linear and the quadratic case and focus on the latter first. We assume that the main contribution to neutrino masses consists of diagrams in which both external neutrinos couple to the respective loop structure via $y_{h_1}$ or $y_{h_2}$.
 Then, the most general neutrino mass matrix reads
\begin{align}\label{amplitude_two_singly_charged}
    U^*m_{\text{diag}}U^\dagger = M_\nu = y^T_{h_1}S_1y_{h_1} + y^T_{h_2}S_2y_{h_2} + y^T_{h_1}Zy_{h_2} + y^T_{h_2}Z^Ty_{h_1}\,,
\end{align}
with the symmetric coupling matrices $S_{1,2}$ and a general matrix $Z$. Multiplying Eq.~(\ref{amplitude_two_singly_charged}) by the respective eigenvectors $v_{h_{1,2}}$ of $y_{h_{1,2}}$ with eigenvalue zero, one obtains three inequivalent complex conditions:
\begin{subequations}
    \begin{align}
    v^T_{h_1}U^*m_{\text{diag}}U^\dagger v_{h_1} & = v^T_{h_1}y^T_{h_2}S_2y_{h_2}v_{h_1},\label{constraint_eigenvector_two_singly_charged1} \\ v^T_{h_2}U^*m_{\text{diag}}U^\dagger v_{h_1} & = v^T_{h_2}y^T_{h_1}Zy_{h_2}v_{h_1}, \label{constraint_eigenvector_two_singly_charged2}\\ v^T_{h_2}U^*m_{\text{diag}}U^\dagger v_{h_2} & = v^T_{h_2}y^T_{h_1}S_1y_{h_1}v_{h_2}\,.
    \label{constraint_eigenvector_two_singly_charged3}\end{align}
\end{subequations}
Contrary to the linear case and the quadratic case discussed for one singly-charged scalar singlet before, here the constraint explicitly involves the matrices $S_{1,2}$ and $Z$ which parametrise the breaking of lepton-number conservation. In that sense, Eqs.~\eqref{constraint_eigenvector_two_singly_charged1}-\eqref{constraint_eigenvector_two_singly_charged3} are in general model-dependent and hence less predictive. In App.~\ref{sec:gzbm} we present one possible neutrino mass model with two singly-charged scalar singlets as an example.

In explicit models, some of the matrices $S_{1,2}, Z$ may vanish in the case of an additional symmetry. Then, the respective expressions above would simplify accordingly and one could solve them as in the linear case.
Still, irrespective of the specific structures in $S_{1,2}$ and $Z$, all Eqs.~\eqref{constraint_eigenvector_two_singly_charged1}-\eqref{constraint_eigenvector_two_singly_charged3} have to be individually satisfied. Thus, the contributions $y^T_{h_1}S_1y_{h_1}$ and $y^T_{h_2}S_2y_{h_2}$ are not independent, but the elements of $y_{h_1}$ and $y_{h_2}$ are intertwined via each of the right-hand sides and in particular also via the left-hand side of Eq.~\eqref{constraint_eigenvector_two_singly_charged2} even if $Z$ is taken to zero.
Only if both $Z$ and one of the matrices $S_1$ and $S_2$ are absent, one trivially recovers the quadratic case for one singly-charged scalar singlet. The same limit is obtained for a large hierarchy between the masses of $h_1$ and $h_2$ since one may integrate out the heavier singlet and attain more predictive power. Lastly, it is straightforward to generalise Eq.~\eqref{amplitude_two_singly_charged} towards the case of $n$ singly-charged scalar singlets generating sizeable contributions to neutrino masses.\footnote{See for instance \cite{Chakrabarty:2018qtt,Nomura:2018vfz} for studies of variants of the Zee-Babu model which contain three singly-charged scalar singlets.}

In the linear case, one would generalise the structure of the neutrino mass matrix towards
\begin{align}
    U^*m_{\text{diag}}U^\dagger = M_\nu = X_1y_{h_1} - y_{h_1}X_1^T + X_2y_{h_2} - y_{h_2}X_2^T\,.
\end{align}
Still, there is no non-trivial limit in which the derived constraints would become independent of the model-dependent physics in $X_{1,2}$.

\section{Conclusions}
\label{sec:conclusions}

We have presented a classification and phenomenological study for scenarios in which a singly-charged scalar singlet particle $h$ generates the main contribution to neutrino masses. Among the SM fermions, $h$ interacts only with the left-handed lepton doublets via an antisymmetric Yukawa coupling $y_h LLh + \text{h.c.}$ at tree level. It is possible to assign charges of lepton number to $h$ and the SM leptons in a way such that it is respected by all renormalisable terms in the Lagrangian. Thus, in order to generate Majorana masses for neutrinos, one needs to introduce a source of lepton-number breaking. Our approach is independent of the details of this breaking. \emph{The only assumption is that the main contribution of neutrino masses is generated by a diagram in which one or both of the external neutrinos couples via $y_h$}.

For the minimal case of just one singlet state, this gives rise to only two possible structures for the neutrino mass matrix. Regarding the Feynman diagram which generates the main contribution to neutrino masses, we distinguish between the ``linear case" in which only one external neutrino is linked to the loop structure via $y_h$, and the ``quadratic case" in which both external neutrinos are.
Several well-known models of neutrino-mass generation fall into those two categories:
The Zee model~\cite{Zee:1980ai,Cheng:1980qt,WOLFENSTEIN198093}  is an example for the linear case and the Zee-Babu~\cite{Zee:1985id,Zee:1985rj,Babu:1988ki} and KNT models~\cite{Krauss:2002px} are examples for the quadratic case.

For each of the cases, we employ the antisymmetry of $y_h$ in flavour space to derive a model-independent constraint which has to be satisfied to guarantee the correct description of the measured mixing and mass hierarchy of neutrinos. In the linear case, the constraint determines two of the magnitudes of the Yukawa coupling matrix elements $|y_h^{ij}|$ in terms of the third one, the two phases of $y_h^{ei}$, $i=\mu,\tau$, neutrino masses, leptonic mixing angles and phases.
In the quadratic case, the two constraints are independent of the neutrino masses and Majorana phases and thus more predictive.

This enables us to perform a phenomenological study applicable to many different types of models. The study is conservative in the sense that no other contributions to the considered observables beyond the ones induced by $h$ are taken into account. If the other new particles involved in a specific model are sufficiently decoupled in the sense that they are very heavy or very weakly coupled to the SM, the phenomenological bounds obtained will approximately coincide with those of the actual model, otherwise the bounds will be weaker.  This is trivially satisfied in an effective field theory framework, in which the singly-charged scalar singlet is much lighter than all other new particles.

For the linear case, the constraint disfavours large hierarchies among the coupling magnitudes $y^{ij}_h$ and hence the available parameter space is mostly shaped by $\mu\to e\gamma$ and $\mu\to 3e$ and other low-energy processes are generally not competitive.
The relative magnitudes $|y^{ij}_h|$ display some sensitivity to the neutrino-mass ordering. Furthermore,
we demonstrate that the region in parameter space where the
Cabibbo Angle Anomaly and the deviation of leptonic gauge couplings from universality, collectively dubbed the ``flavour anomalies", are explained by $h$ is compatible with, albeit not preferred by neutrino masses. A conclusive measurement of $M_W > \SI{80.35}{\giga\electronvolt}$ would imply that $h$ cannot explain the flavour anomalies.

For the quadratic case, the parameter space is strongly constrained by neutrino masses and thus the scenario is very predictive. The leptonic gauge couplings do not receive large contributions and thus it is not possible to explain their deviation from universality as indicated by current data and neither the Cabibbo Angle Anomaly. Furthermore, there is a tight correlation between the different radiative charged-lepton decays and hence any signal of a radiative flavour-violating tau decay at Belle II would imply that low-energy effects of new physics cannot be assumed to be dominated by $h$. Also, one may derive sharp predictions for the branching ratios of the different decay  channels to satisfy $\text{Br}(h\to\tau\nu)\simeq\text{Br}(h\to\mu\nu)\sim\num{0.4}$ and $\text{Br}(h\to e\nu)\lesssim\num{0.2}$ for NO and $\text{Br}(h\to\tau\nu)\sim\num{0.3}$, $\text{Br}(h\to\mu\nu)\sim\num{0.2}$ and $\text{Br}(h\to e\nu)\sim 0.5$ for IO. The branching ratios $\text{Br}(h\to\tau\nu)$ and $\text{Br}(h\to\mu\nu$) exhibit a slight dependence on the Dirac CP phase $\delta$ for NO.

Finally, we commented on the generalisation of our framework to multiple singly-charged scalar singlets. One may also derive constraints in that case, but they depend on the breaking of lepton number and thus do not allow for a model-independent study.

To conclude, this study of the singly-charged scalar singlet $h$ is a neat example of a model-independent approach towards neutrino masses and their phenomenological implications. The constraints originate only from the form of the neutrino mass matrix and the antisymmetry of the Yukawa coupling of $h$ to left-handed lepton doublets. We leave the discussion of other simplified neutrino mass scenarios for future work.

\begin{acknowledgments}
TF and MS acknowledge support by the Australian Research Council via the Discovery Project grant DP200101470. JHG is supported by the Generalitat Valenciana through the GenT Excellence Program (CIDEGENT/2020/020). We acknowledge the use of \texttt{python-ternary}~\cite{pythonternary}.
\end{acknowledgments}

\appendix

\section{Neutrino-Mass Constraint Spelt in Full}
\label{sec:nmc}

Written out, the constraint in Eq.~(\ref{constraint_eigenvector}) explicitly reads
\begin{align}
    & \Bigg(\Big(c_{13}^2 m_3 c_{23}^2+e^{-2 i \eta _2} m_2 \left(e^{-i \delta } c_{23} s_{12} s_{13}+c_{12} s_{23}\right)^2 + \\
    & e^{-2 i \eta _1} m_1 \left(e^{-i \delta } c_{12} c_{23} s_{13}-s_{12} s_{23}\right)^2\Big) y_h^{\text{e$\mu $}} \nonumber\\
    - & \Big(c_{23} m_3 s_{23} c_{13}^2+e^{-2 i \eta _1} m_1 \left(e^{-i \delta } c_{12} c_{23} s_{13}-s_{12} s_{23}\right) \left(c_{23} s_{12}+e^{-i \delta } c_{12} s_{13} s_{23}\right) \nonumber\\
    - & e^{-2 i \eta _2} m_2 \left(e^{-i \delta } c_{23} s_{12} s_{13}+c_{12} s_{23}\right) \left(c_{12} c_{23}-e^{-i \delta } s_{12} s_{13} s_{23}\right)\Big) y_h^{\text{e$\tau $}} \nonumber\\
    + & c_{13} \Big(e^{i \delta } c_{23} m_3 s_{13}-e^{-2 i \eta _2} m_2 s_{12} \left(e^{-i \delta } c_{23} s_{12} s_{13}+c_{12} s_{23}\right) \nonumber\\
    + & e^{-2 i \eta _1} c_{12} m_1 \left(s_{12} s_{23}-e^{-i \delta } c_{12} c_{23} s_{13}\right)\Big) y_h^{\mu \tau }\Bigg) y_h^{\text{e$\mu $}} \nonumber\\
    - & \Bigg(\Big(c_{23} m_3 s_{23} c_{13}^2+e^{-2 i \eta _1} m_1 \left(e^{-i \delta } c_{12} c_{23} s_{13}-s_{12} s_{23}\right) \left(c_{23} s_{12}+e^{-i \delta } c_{12} s_{13} s_{23}\right) \nonumber\\
    - & e^{-2 i \eta _2} m_2 \left(e^{-i \delta } c_{23} s_{12} s_{13}+c_{12} s_{23}\right) \left(c_{12} c_{23}-e^{-i \delta } s_{12} s_{13} s_{23}\right)\Big) y_h^{\text{e$\mu $}} \nonumber\\
    - & \Big(c_{13}^2 m_3 s_{23}^2+e^{-2 i \eta _1} m_1 \left(c_{23} s_{12}+e^{-i \delta } c_{12} s_{13} s_{23}\right)^2 \nonumber\\
    & + e^{-2 i \eta _2} m_2 \left(c_{12} c_{23}-e^{-i \delta } s_{12} s_{13} s_{23}\right)^2\Big) y_h^{\text{e$\tau $}} \nonumber\\
    - & e^{i \delta } c_{13} \Big(e^{-2 i \left(\delta +\eta _1\right)} m_1 s_{13} s_{23} c_{12}^2+e^{-i \delta } c_{23} \left(e^{-2 i \eta _1} m_1-e^{-2 i \eta _2} m_2\right) s_{12} c_{12} \nonumber\\
    - & \left(m_3-e^{-2 i \left(\delta +\eta _2\right)} m_2 s_{12}^2\right) s_{13} s_{23}\Big) y_h^{\mu \tau }\Bigg) y_h^{\text{e$\tau $}} \nonumber\\
    + & \Bigg(c_{13} \Big(e^{i \delta } c_{23} m_3 s_{13}-e^{-2 i \eta _2} m_2 s_{12} \left(e^{-i \delta } c_{23} s_{12} s_{13}+c_{12} s_{23}\right) \nonumber\\
    + & e^{-2 i \eta _1} c_{12} m_1 \left(s_{12} s_{23}-e^{-i \delta } c_{12} c_{23} s_{13}\right)\Big) y_h^{\text{e$\mu $}} \nonumber\\
    + & e^{i \delta } c_{13} \Big(e^{-2 i \left(\delta +\eta _1\right)} m_1 s_{13} s_{23} c_{12}^2+e^{-i \delta } c_{23} \left(e^{-2 i \eta _1} m_1-e^{-2 i \eta _2} m_2\right) s_{12} c_{12} \nonumber\\
    -& \left(m_3-e^{-2 i \left(\delta +\eta _2\right)} m_2 s_{12}^2\right) s_{13} s_{23}\Big) y_h^{\text{e$\tau $}} \nonumber\\\nonumber
    + & \left(e^{-2 i \eta _2} m_2 s_{12}^2 c_{13}^2+e^{-2 i \eta _1} c_{12}^2 m_1 c_{13}^2+e^{2 i \delta } m_3 s_{13}^2\right) y_h^{\mu \tau }\Bigg) y_h^{\mu \tau } = 0\,,
\end{align}
with the abbreviations $s_{ij} \equiv \sin(\theta_{ij})$ and $c_{ij} \equiv \cos(\theta_{ij})$.

\section{Effective Four-Lepton Operator}
\label{sec:effOp}
Starting from the full theory as defined via Eqs.~\eqref{kinetic_Lagrangian} and \eqref{leptonic_Lagrangian}, one obtains the lowest-order solution to the classical equation of motion for $h$:
\begin{equation}
    h = -\frac{(y^{ij}_h)^*}{M^2_h}L^\dagger_j L^\dagger_i\,.
\end{equation}
The resulting effective Lagrangian reads
\begin{equation}
    \mathcal{L}_{\text{eff}} = \frac{(y^{ij}_h)
    ^*y^{kl}_h}{M^2_h}L^\dagger_jL^\dagger_iL_kL_l = \frac{(y^{ik}_h)^*y^{jl}_h}{M^2_h}L^{\dagger\alpha}_i\bar\sigma^\mu L_{j\alpha}L^{\dagger\beta}_k\bar\sigma_\mu L_{l\beta}.
\end{equation}
In order to derive this result, one first observes that
\begin{align}
    L^\dagger_jL^\dagger_iL_kL_l = L^{\dagger\alpha}_jL^{\dagger\beta}_iL_{k\beta}L_{l\alpha} - L^{\dagger\alpha}_jL^{\dagger\beta}_iL_{k\alpha}L_{l\beta}.
\end{align}
Together with the antisymmetry of $y_h$, this implies
\begin{align}
    \frac{(y^{ij}_h)
    ^*y^{kl}_h}{M^2_h}L^\dagger_jL^\dagger_iL_kL_l
    = 2\frac{(y^{ij}_h)^*y^{kl}_h}{M^2_h}L^{\dagger\alpha}_jL^{\dagger\beta}_iL_{k\beta}L_{l\alpha}.
\end{align}
Then, applying a Fierz transformation and relabeling flavour indices yields the result
\begin{align}
    \frac{(y^{ij}_h)
    ^*y^{kl}_h}{M^2_h}L^\dagger_jL^\dagger_iL_kL_l & = \frac{(y^{ij}_h)^*y^{kl}_h}{M^2_h}L^{\dagger\alpha}_j\bar\sigma^\mu L_{l\alpha}L^{\dagger\beta}_i\bar\sigma_\mu L_{k\beta} \\\nonumber
    & = \frac{(y^{ik}_h)^*y^{jl}_h}{M^2_h}L^{\dagger\alpha}_i\bar\sigma^\mu L_{j\alpha}L^{\dagger\beta}_k\bar\sigma_\mu L_{l\beta} \equiv C^{ijkl}_{LL}\mathcal{O}_{LL,ijkl}
    .
\end{align}

\section{$\mu-e$ Conversion in Nuclei}
\label{sec:Mu2E}
We consider the photon-penguin contribution to the effective Lagrangian for $\mu-e$ conversion in nuclei and neglect all other contributions following \cite{Crivellin:2020klg}. In addition to the short-range contribution which has been discussed in \cite{Crivellin:2020klg} we also include the relevant long-range contribution. Following \cite{Kitano:2002mt}, we identify the relevant terms
\begin{align}
\mathcal{L}_{\text{eff}} & = -4\sqrt{2} G_F \,\left(m_\mu A_R \bar \mu \sigma^{\mu\nu} e F_{\mu\nu}
+\mathrm{h.c.}\right)
-\frac{G_F}{\sqrt{2}}\sum_{q=u,d,s}\Big[
g_{LV(q)}
\,
e^\dagger  \bar\sigma^\mu \mu
\,(q^\dagger \bar\sigma_\mu q + \bar q \sigma_\mu \bar q^\dagger)
\Big]
\end{align}
in the effective Lagrangian, where the ``barred" fields denote the charge-conjugates of the respective right-handed fields and we employ the 2-component notation for spinors as detailed in~\cite{Dreiner:2008tw}. The Wilson coefficients are given by~\cite{Crivellin:2020klg}
\begin{align}
    A_R & = -\frac{1}{2\sqrt{2}G_F} \frac{\sqrt{4\pi\alpha_{\text{EM}}}}{96\pi^2 M_h^2} y^{e\tau}_h(y^{\mu\tau}_h)^*,
    &
    g_{LV(q)} & = -\frac{\sqrt{2}}{G_F} \frac{4\pi\alpha_{\text{EM}}Q_q}{72\pi^2 M_h^2} (y^{e\tau}_h)^*y^{\mu\tau}_h\,,
\end{align}
where $Q_q$ denotes the electric quark charge, $Q_u=\tfrac23$ and $Q_d=-\tfrac13$.
The resulting conversion rate is~\cite{Kitano:2002mt}
\begin{align}
    \omega_{\rm conv} & = 2G_F^2 m_\mu^5 \left| A_R^* D + \tilde g_{LV}^{(p)} V^{(p)} + \tilde g_{LV}^{(n)} V^{(n)}\right|^2
\end{align}
in terms of the couplings to protons and neutrons. The coupling to neutrons vanishes, $\tilde g_{LV}^{(n)}  =  g_{LV(u)} +2 g_{LV(d)} =0$, because the photon-penguin contribution is proportional to the electric charge of the nucleon, and the effective coupling to the proton is
\begin{align}
    \tilde g_{LV}^{(p)} & = 2 g_{LV(u)} + g_{LV(d)}
    =-\frac{\sqrt{2}\alpha_{\text{EM}}}{18\pi G_F M_h^2} (y^{e\tau}_h)^*y^{\mu\tau}_h.
\end{align}
Hence we find the conversion rate
\begin{align}
    \omega_{\rm conv} & = \left|(y^{e\tau}_h)^* y^{\mu\tau}_h\right|^2 \left| \frac{\alpha_{\rm EM}^{1/2} D}{96\pi^{3/2}}  + \frac{\alpha_{\rm EM} V^{(p)}}{9\pi} \right|^2 \frac{m_\mu^5}{M_h^4}
    \;.
\end{align}
The experimental limits for $\mu-e$ conversion are generally quoted in terms of the ratio of the conversion rate $\omega_{\rm conv}$ over the capture rate $\omega_{\rm capt}$~\cite{Suzuki:1987jf, Kitano:2002mt}, $\text{Br}(\mu\to e\text{;}\;\text{X}) \equiv \omega^\text{X}_{\text{conv}}/\omega^\text{X}_{\text{capt}}$. For $X = \text{Au},\text{Al},\text{Ti}$ we use
\begin{align}
    \omega^\text{Au}_{\rm capt} & = \SI{13.06E6}{\per\second},
    &
    \omega^\text{Al}_{\rm capt} & = \SI{0.7054E6}{\per\second},
    &
    \omega^\text{Ti}_{\rm capt} & = \SI{2.59E6}{\per\second}.
\end{align}
Currently, the SINDRUM II experiment places the strongest limit on $\mu-e$ conversion in gold~\cite{Bertl:2006up}
with
$\text{Br}(\mu\to e\text{;}\;\text{Au}) \equiv \omega^\text{Au}_{\text{conv}}/\omega^\text{Au}_{\text{capt}} < \num{7E-13}$. In the coming years, several experiments with improved sensitivity will probe unexplored parameter space using $\mu-e$ conversion: The Mu2e experiment at Fermilab~\cite{Kutschke:2011ux} and the COMET experiment~\cite{Hungerford:2009zz} are expected to reach a sensitivity of $\num{6E-17}$ and $\num{2.6E-17}$, respectively, for an aluminum target. Ultimately, PRISM/PRIME~\cite{Kuno:2005mm} is projected to reach a sensitivity of $10^{-18}$ for a titanium target.
The relevant overlap integrals for the long-range and short-range photon-penguin contributions to $\mu-e$ conversion in gold, aluminum and titanium are given by $D$, $V^{(n)}$ and $V^{(p)}$:
\begin{align}
D_{\rm Au} & = 0.189\;,
&
    V_{\rm Au}^{(p)} & =
    0.0974\;,
    &
    V_{\rm Au}^{(n)} & = 0.146\;,
    \nonumber\\
    D_{\rm Al} & = 0.0362\;,
&
    V_{\rm Al}^{(p)} & =
    0.0161\;,
    &
    V_{\rm Al}^{(n)} & = 0.0173\;,
 \\\nonumber
    D_{\rm Ti} & = 0.0864\;,
&
    V_{\rm Ti}^{(p)} & =
    0.0396\;,
    &
    V_{\rm Ti}^{(n)} & = 0.0468
    \;.
\end{align}

\section{Generalised Zee-Babu Model}
\label{sec:gzbm}
A natural example of the quadratic case with two singly-charged scalar singlets is given by a generalised version of the Zee-Babu model. Hence, consider the extension of the SM particle content by two singly-charged scalar singlets $h_1$ and $h_2$ and a doubly-charged scalar singlet $k$. Assuming the mass basis for the singly-charged scalar singlets and neglecting all terms in the scalar potential which are unrelated to the breaking of lepton-number conservation, one finds the following Lagrangian:
\begin{align}
    \mathcal{L} & = -h^*_1(D^\mu D_\mu + M^2_1)h_1 - h^*_2(D^\mu D_\mu + M^2_2)h_2 - k^*(D^\mu D_\mu + M^2_k)k \nonumber\\
    & \quad - \left(\left(\mu_1h^2_1 + \mu_2h^2_2 + \mu_{12}h_1h_2\right)k^* + \text{h.c.}\right) \\\nonumber
    & \quad - \left(y^*_eL^\dagger H\bar e^\dagger + y_{h_1}LLh_1 + y_{h_2}LLh_2 + y_k\bar e^\dagger\bar e^\dagger k + \text{h.c.}\right)\,.
\end{align}
The contribution to neutrino masses corresponding to $S_1$ ($S_2$) which is defined in Eq.~\eqref{amplitude_two_singly_charged} can be obtained as in the Zee-Babu model and is proportional to $\mu_1/M^2_k$ ($\mu_2/M^2_k$) in the limit $M_k/M_h \gg 1$.\footnote{This assumption may be relaxed without altering the form of the dominant contribution to neutrino masses. Still, the relevant loop function acquires a simple form only in the limits $M_k/M_h \gg 1$ and $M_k/M_h \to 0$~\cite{McDonald:2003zj,Nebot:2007bc,Herrero-Garcia:2014hfa}.} To our knowledge, a similar limit for the contribution corresponding to $Z$ has not been considered yet, but in analogy it may be expected to be proportional to $\mu_{12}/M^2_k$. There are up to fifteen parameters in the model which are directly linked to neutrino masses: $\mu_1$, $\mu_2$, $\mu_{12}$, six couplings in $y_k$ and three couplings both in $y_{h_1}$ and in $y_{h_2}$. In general, up to six of them can be determined via the constraint in Eqs.~(\ref{constraint_eigenvector_two_singly_charged1})-(\ref{constraint_eigenvector_two_singly_charged3}) upon fixing the other parameters. A detailed study of the full generalised Zee-Babu model is left for future work.

In the case of a further hierarchy between the masses $M_1$ and $M_2$ (which is assumed not to be cancelled by another hierarchy in $|\mu_1|$ and $|\mu_2|$), such that the contribution of the heavier singly-charged scalar singlet both to neutrino masses and to other flavour observables can be neglected with respect to the lighter one, one may integrate out the former. Hence, assuming $M_2 \gg M_{1,k}$ and integrating out $h_2$, we obtain the following effective Lagrangian up to dimension-6 terms:
\begin{align}
    \mathcal{L}_{\text{eff}} & \supseteq -h^*_1(D^\mu D_\mu + M^2_1)h_1 - k^*(D^\mu D_\mu + M^2_k)k - \left(\mu_1h^2_1k^* + \text{h.c.}\right) - \left(y^{ij}_{h_1}L_iL_jh_1 + \text{h.c.}\right) \nonumber\\
    & \quad - \left(\frac{\mu^*_2y^{ij}_{h_2}y^{kl}_{h_2}}{M^4_2}kL_iL_jL_kL_l + \text{h.c.}\right) + \left(\frac{\mu^*_{12}y^{ij}_{h_2}}{M^2_2}kh^*_1L_iL_j + \text{h.c.}\right) + \frac{(y^{ij}_{h_2})^*y^{kl}_{h_2}}{M^2_2}L_kL_lL^\dagger_jL^\dagger_i \nonumber\\
    & \quad - \left(y_k\bar e^\dagger\bar e^\dagger k + \text{h.c.}\right)\,.
\end{align}
This corresponds to the Zee-Babu model extended by effective interactions. Thus, in the effective field theory limit $M_2\gg M_{1,k}$ the dominant contribution to neutrino masses is given by the renormalisable terms and therefore the same as in Eq.~\eqref{eq:SZB} for $h\to h_1$.

\bibliography{refs}
\end{document}